\documentclass[aps,reprint,onecolumn,notitlepage]{revtex4-1}
\usepackage{amsmath, amssymb, color, enumitem, graphicx}
\usepackage{natbib}

\newcommand{\rey}{\text{Re}}
\newcommand{\pe}{\text{Pe}}
\newcommand{\ca}{\text{Ca}}
\newcommand{\ei}{\text{Ei}}

\begin{abstract}
Experimental observations indicate that chemically active droplets suspended in a surfactant-laden fluid can self-propel spontaneously. The onset of this motion is attributed to a symmetry-breaking Marangoni instability resulting from the nonlinear advective coupling of the distribution of surfactant to the hydrodynamic flow generated by Marangoni stresses at the droplet's surface. Here, we use weakly nonlinear analysis to characterize the self-propulsion near the instability threshold and the influence of the droplet's deformability. We report that in vicinity of the threshold, deformability enhances self-propulsion of viscous droplets, but hinders propulsion of drops that are roughly less viscous than the surrounding fluid. Our asymptotics further reveals that droplet deformability may alter the type of bifurcation leading to symmetry breaking: for moderately deformable droplets the onset of self-propulsion is transcritical and a regime of steady self-propulsion is stable; while in the case of highly deformable drops, no steady flows can be found within the asymptotic limit considered in this paper suggesting that the bifurcation is subcritical.
\end{abstract}

\begin{document}


\title{Self-propulsion near the onset of Marangoni instability of deformable active droplets}
\author{Matvey Morozov}
\author{S{\'e}bastien Michelin}
\email{sebastien.michelin@ladhyx.polytechnique.fr}
\affiliation{LadHyX -- D{\'e}partement de M{\'e}canique, 
  {\'E}cole Polytechnique -- CNRS, 91128 Palaiseau Cedex, France}

\maketitle


\section{Introduction}
Several experimental studies have recently reported self-propulsion of active droplets, whose swimming motion in viscous flows arise from spontaneously-generated surface tension gradients~\cite{Maass16, Ryazantsev17}. These active droplets can  rely either on chemical reactions~\cite{Thutupalli11} or solubilization~\cite{Izri14, Moerman17, Kruger16} as a source of chemical energy to power their self-propulsion for many hours at velocities up to one diameter per second. Experimental observations further indicate that self-propelling microdroplets may exhibit complex dynamical behaviour including straight, curved or chaotic trajectories~\cite{Suga18}. Recent studies have focused specifically on the type of chemical activity~\cite{Herminghaus14, Izri14, Nagasaka17}, the physical properties of the fluid making up the drop~\cite{Kruger16}, the presence of other active droplets~\cite{Maass16, Moerman17} or geometrical constraints on the droplet's environment~\cite{Kruger16b,Jin18}. Sophisticated dynamics paired with potential biocompatibility makes active droplets a prime candidate for modeling and engineering of biological systems~\cite{Izri14, Maass16, Nagasaka17}, as well as for studying and characterization of collective motion of self-propelled agents.

In order to elucidate the mechanisms responsible for their self-propulsion, their complex individual motion and their interactions, active droplets have also attracted much theoretical and modelling effort. Unlike other currently popular microswimmers, such as bacteria or Janus particles, active droplets do not possess inherent asymmetry and, thus, rely on a symmetry-breaking instability to initiate self-propulsion~\citep{Herminghaus14, Ryazantsev17, Yoshinaga17}. In a typical scenario, the instability establishes a concentration or temperature gradient, which produces an uneven stress distribution at the droplet interface, and the droplet may self-propel due to the Marangoni effect.
Naturally, spontaneous loss of symmetry calls for a bifurcation analysis: Rednikov~{\it et al.} developed a weakly nonlinear theory of a self-propelling nondeformable active droplet in the presence of buoyancy force~\cite{Rednikov94, Rednikov94b}. In the limit of small P{\'e}clet and Reynolds numbers, Rednikov~{\it et al.} showed that the balance of self-propulsion and buoyancy force spawns multiple regimes of steady propulsion of the droplet featuring different flow patterns within and outside of the propelling drop. In contrast, recent theoretical works assume a finite value of P{\'e}clet number to emphasize the role of advection in the symmetry breaking instability enabling the transport of active droplets; several models sharing the same basic ingredients have been considered, which differ on the exact production mechanism, transport or bulk reactivity of the chemical solute responsible for the Marangoni flows at the heart of the symmetry-breaking instability~\cite{Thutupalli11, Yabunaka12, Yoshinaga12, Izri14, Moerman17}. In particular, Yoshinaga adopted weakly nonlinear approach to the problem and derived amplitude equations governing the droplet dynamics near the onset of self-propulsion in the presence of a linear chemical reaction in the bulk fluid~\cite{Yoshinaga12}.

Three-dimensional Marangoni flow stirred by an active droplet was considered by Schmitt and Stark, who employed their results to engineer a setup for guiding active drops with laser light~\cite{Schmitt16}. Dynamics of active drops can be also modeled based on reaction-diffusion equations~\citep{Shitara11, Schmitt13}. In particular, Shitara~\textit{et al.} investigated the motion for an isolated domain confined in an excitable reaction-diffusion system and demonstrated that there are three basic motions of the domain: straight motion, rotating motion, and helical motion~\cite{Shitara11}. Diffusion-advection-reaction equation-based model developed by Schmitt and Stark also yields several dynamical regimes: depending on the strength of the Marangoni effect, the droplet may self-propel steadily, spontaneously stop, or oscillate~\cite{Schmitt13}. The effect of chemical product that changes the interfacial energy of a droplet and thus affects the symmetry-breaking Marangoni instability was investigated by Yabunaka~\textit{et al.}~\cite{Yabunaka12}, while Yabunaka and Yoshinaga recently analysed the hydrodynamic and chemical interactions of two droplets, and their resulting collision dynamics~\cite{Yabunaka16}. The interested reader is referred to the recent reviews of Herminghaus~\textit{et al.}~\cite{Herminghaus14} and Maass~\textit{et al.}~\cite{Maass16} on the self-propulsion of active droplets for a more exhaustive review of both experimental and theoretical work on this topic.

It should be noted that the transport due to Marangoni forces is not exclusive to submerged droplets: the same mobility mechanism applies to swimmers moving along a liquid surface~\cite{Wurger14, Frenkel18}. Self-propulsion enabled by a symmetry-breaking instability was also observed in active particles driven by diffusiophoresis~\cite{Moran17}. In particular, Michelin~\textit{et al.} have demonstrated theoretically that the flow around a chemically-active spherical autophoretic particle may lose its stability via symmetry-breaking bifurcation resulting in self-propulsion of the particle~\cite{Michelin13}. Together with the chemical activity of the particle or droplet, the advective transport of chemical species by the flow field they generate through Marangoni stresses or phoretic slip velocities is, thus, the key ingredient leading to self-propulsion beyond a certain threshold required to overcome the effect of diffusion. Even below the critical threshold for propulsion, chemical activity and phoretic mobility of the particles were also shown to significantly impact their response to outer flows (e.g., phoretic drag reduction~\cite{Yariv17}).

Self-propelled droplets and rigid diffusio-phoretic particles share many similarities but differ on one key feature, namely the origin of the flow field in response to a concentration gradient: for rigid phoretic particles, the flow stems from nonzero slip velocity at the particle surface in response to a chemical gradients, whereas mobility of Marangoni droplets is sustained by interfacial stresses~\citep{Moran17, Yoshinaga17}. It should be noted that both mechanisms can be described within the same framework, and in fact coexist in the case of droplets, although phoretic effects are essentially negligible in front of Marangoni forcing except for very viscous droplets~\citep{Anderson89}.

As noted in the opening paragraph, physics of spontaneous self-propulsion is complex and represents considerable interest. In particular, recent experimental observations of liquid crystal droplets revealed the coupling between director field inside the drop and the trajectory of droplet self-propulsion~\citep{Kruger16}. Importance of the internal droplet structure was further investigated by Kree~\textit{et al.}, who developed a theoretical model of self-propulsion of a spherical droplet containing a rigid skeleton~\cite{Kree17}. Even in the absence of advection, the geometry of active particles was also shown recently to strongly affect or control the direction and magnitude of propulsion as well as their hydrodynamic signature~\cite{Lauga16, Nourhani16, Michelin17, Ibrahim18}. Shape can also act as a symmetry-breaking mechanism for chemically-homogeneous systems~\cite{Shklyaev14, Michelin15}. Although their Laplace pressure remains typically greater than the hydrodynamic viscous stresses they sustain from the surrounding fluid, self-propelled droplets do not have a fixed shape but may deform under the effect of surfactant gradients or fluid motion. One of the present paper's main objectives is to characterize the fundamental effect of surface deformability on mode competition and self-propulsion characteristics of active droplets. Deformability typically accompanies self-propulsion of microorganisms and active particles in general~\cite{Winklbauer15, Ohta17}. Dynamics of deformable droplets driven by the Marangoni effect was recently investigated theoretically by Yoshinaga~\cite{Yoshinaga14} and by means of lattice-Boltzmann simulations by Fadda~\textit{et al.}~\cite{Fadda17}.

A large number of microscopic active droplets sustained in a bulk liquid constitute an active emulsion. It has been established that collective behavior of drops in active emulsions may follow several distinct scenarios of symmetry breaking~\cite{Herminghaus14}. The focus of the present paper is on the self-propulsion of a single active droplet, and such collective phenomena are beyond the scope of the present paper; we refer the reader interested in the theory of active emulsions to the recent review by Weber~\textit{et al.}~\cite{Weber18}.

Active droplets have typical diameters of a few tens of $\mu$m, and swim at a few $\mu$m.s$^{-1}$. Hence, viscous stresses are typically much larger than inertial forces. Following recent theoretical analyses of active droplets, see Refs.~\cite{Izri14, Yoshinaga14, Herminghaus14, Maass16}, we consider the droplet dynamics within the framework of Stokes flows and for moderate values of the P{\'e}clet number. We note that the self-propulsion of an active deformable droplet in the limit of high solutal P{\'e}clet number, $\pe \gg 1$, and vanishing thermal P{\'e}clet number was investigated by Golovin~\textit{et al.}~\cite{Golovin89}. Unlike Yabunaka~\textit{et al.}~\cite{Yabunaka12} and Yoshinaga~\textit{et al.}~\cite{Yoshinaga12}, we disregard any chemical reaction in the bulk fluid both within and outside the droplet. That is, in our model activity is sustained by a reaction at the droplet interface which roughly corresponds to the micellar dissolution under moderate surfactant concentration (i.e., lower than the critical micelle concentration), as observed by Moerman~\textit{et al.}~\cite{Moerman17}. We adopt an asymptotic approach to the problem at hand to obtain the self-propulsion characteristics near the onset of propulsion as well as analyze the stability of the steady state solutions. As opposed to analyses of Rednikov~\textit{et al.}~\cite{Rednikov94, Rednikov94b}, we include dynamic deformability of the droplet interface into consideration and build our argument based on both the investigation of steady states and explicit stability analysis of these states. The latter allows us to distinguish between physically different temporal scales involved in the onset of Marangoni instability, thus providing additional insight into the competition of different physical mechanisms driving the droplet dynamics.

The paper is organized as follows. In \S~\ref{statement}, the mathematical formulation of the problem is outlined and relevant dimensionless parameters are defined. Neutrally stable eigenmodes of the linearized problem are obtained in \S~\ref{neutral_modes}. \S~\ref{weakly_nonlin} presents a weakly nonlinear analysis of the problem in order to identify and characterize the steady flow regimes emerging due to saturation of neutrally stable modes above the instability threshold. In \S~\ref{lin1}, linear stability analysis of these steady states is discussed and is employed to estimate the typical time scales associated with saturation of different instability modes. Finally, we discuss our findings in \S~\ref{discussion} and present some perspectives.

\section{Physical problem and model}
\label{statement}
The focus of the present work is the spontaneous propulsion and fluid motion generated by active droplets under the effect of the Marangoni instability, and more specifically the effect of interface deformability on the dynamics of an active droplet near the onset of self-propulsion.  We focus on an axisymmetric problem and employ spherical polar coordinates $(r,\mu=\cos\theta)$ centered at the droplet's center of mass.

\subsection{Governing equations and boundary conditions}
\begin{figure}
\centering
  \includegraphics[scale=0.75]{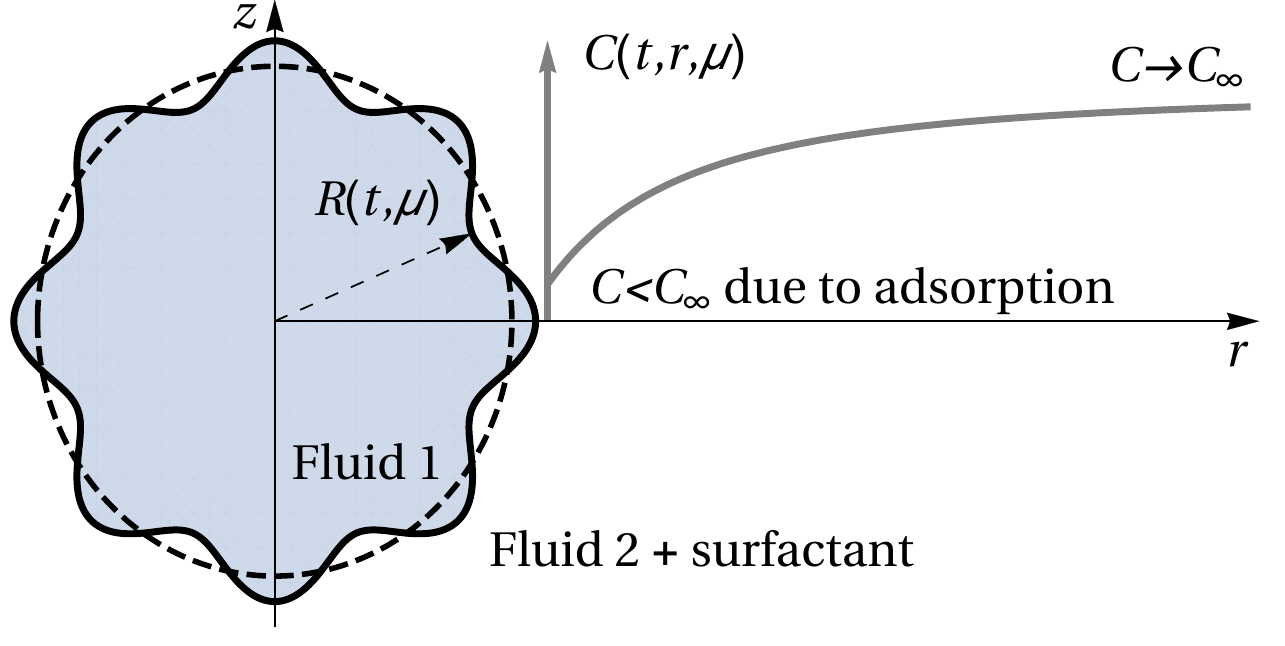}
  \caption{
  Cross section of an axisymmetric deformable active droplet undergoing gradual micellar dissolution. As the droplet dissolves, it adsorbs surfactant that powers the chemical reaction at the droplet interface, thus sustaining the dissolution.
  }
  \label{drop_sketch}
\end{figure}

A liquid droplet of a Newtonian fluid is considered here, with density $\rho_i$ and dynamic viscosity $\eta_i$, submerged in a second Newtonian fluid of density $\rho_o$ and viscosity $\eta_o$, containing a surfactant solute of concentration $C$, as sketched in figure~\ref{drop_sketch}. Note that different experimental setups employ different inner and outer fluids: for instance, Izri~\textit{et al.} observed water droplets in oil~\cite{Izri14}, while others used oil droplets in water~\cite{Moerman17}. To remain general, we denote in the following by subscripts $i$ and $o$ the quantities relevant to the inner and outer fluid, respectively. Far from the droplet, the concentration of surfactant molecules is $\mathcal{C}_\infty$. In the following, we consider axisymmetric deformations of the droplet under flow and Marangoni stresses, whose surface is thus described in spherical polar coordinates by $r=R(t,\mu)$ with $\mu=\cos\theta$. Noting $R_0$ the radius of the drop at rest (i.e., when it is spherical), the radius of the deformed droplet writes
\begin{equation}
  \label{shape}
  R(t,\mu) = R_0 ( 1 + \xi(t,\mu) ).
\end{equation}
Naturally, the presence of deformations would not only contribute to the position of the droplet interface, but also to its curvature. That is, normal and tangential vectors to the interface are now functions of $t$ and $\mu$, thus affecting the boundary conditions formulated below.

Multiple physico-chemical mechanisms have been identified in experiments leading to the self-propulsion of active droplets~\cite{Herminghaus14}. In the following, we explicitly refer to the molecular pathway identified in the experiments ofMoerman~\textit{et al.}~\cite{Moerman17}. Yet the formalism presented here is completely general and could easily be applied to the micellar pathway relevant to the experiments of Izri~\textit{et al.}~\cite{Izri14}. In the molecular pathway leading to the solubilization of the oil phase into the aqueous solution, surfactant molecules are absorbed at the surface and swollen micelles are released leading to slow decrease of the droplet size. In experiments, typical time of droplet dissolution is substantially longer than the time scale associated with self-propulsion~\cite{Izri14, Moerman17}, so that this dissolution process can be neglected and the volume of the droplet is assumed constant, $V = const$, while the droplet consumes surfactant molecules at a fixed rate ${\cal A} > 0$,
\begin{equation}
  \label{c_consump}
  {\cal D} \textbf{n} \cdot \nabla C = {\cal A}
  \qquad \text{at  } r = R,
\end{equation}
where $\textbf{n}$ is the outward normal to the droplet interface. Surfactant molecules do not penetrate into the droplet. Thus, advection-diffusion of surfactant should only be taken into account outside of the drop,
\begin{equation}
  \label{eqs_ad}
  \partial_t C + \textbf{u}_o \cdot \nabla C = {\cal D} \nabla^2 C,
\end{equation}
where $\partial_t$ denotes the partial derivative with respect to time $t$, $\textbf{u}_o$ is the flow velocity outside of the drop, and ${\cal D}$ denotes molecular diffusivity of the surfactant in the outer fluid. Naturally, far away from the droplet surfactant concentration reaches a constant value,
\begin{equation}
  \label{far_c}
  C \rightarrow {\cal C}_\infty
  \qquad \textrm{for  } r \rightarrow \infty.
\end{equation}

The presence of surfactants at the droplet's surface modifies its interfacial tension $\gamma$. Assuming that adsorption/desorption of surfactant molecules at the fluid-fluid interface occurs instantaneously compared to its transport in the outer fluid, concentration of the adsorbed surfactants is $\propto \left. C \right|_{r=R}$~\cite{Baret69}. We further linearize the relationship between $\gamma$ and $\left. C \right|_{r=R}$, 
\begin{equation}
  \gamma = \gamma_0 - \gamma_C \left( \left. C \right|_{r=R} 
    - {\cal C}_\infty + {\cal A} R_0 / {\cal D} \right),
\end{equation}
and note ${\gamma_C \equiv \left. -\mathrm{d}\gamma / \mathrm{d}C \right|_{C = {\cal C}_\infty - {\cal A} R_0 / {\cal D}}> 0}$ with ${{\cal C}_\infty - {\cal A} R_0 / {\cal D}}$ corresponding to the surfactant concentration at $r=R$ in the absence of flow. The stress balance at the interface then writes in vector form (i.e., accounting for both normal and tangential stresses)
\begin{equation}
  \label{stress_bc}
  ( \boldsymbol \sigma_o - \boldsymbol \sigma_i ) \cdot \mathbf{n}
  + \nabla \cdot [ ( \mathbf{I} - \mathbf{nn} ) \gamma ] 
  = 0 \qquad \textrm{at  } r = R,
\end{equation}
where $\mathbf{I}$ is the identity tensor, $\boldsymbol\sigma_{i,o}=-P_{i,o}\mathbf{I}+\boldsymbol\tau_{i,o}$ is the hydrodynamic stress tensor with $\boldsymbol\tau_{i,o}=\eta_{i,o}(\nabla\mathbf{u}_{i,o}+\nabla\mathbf{u}_{i,o}^T)$ its viscous part. The continuity of the fluid's velocity and impermeability of the droplet's surface is written as 
\begin{equation}
  \label{kin_cont}
  R_0 \frac{\partial \xi}{\partial t}
  - R_0 \frac{ \sqrt{1-\mu^2} }{r}
      \frac{\partial \xi}{\partial \mu} (\textbf{u}_o \cdot \textbf{e}_\theta)
  = \textbf{u}_o \cdot \textbf{e}_r,
  \qquad \textbf{u}_o = \textbf{u}_i 
  \qquad \textrm{at  } r = R.
\end{equation}

In experiments, typical droplet sizes and velocities are $R_0\sim 10$~$\mu$m and $V\sim 10\mu$~m.s$^{-1}$, respectively~\cite{Izri14, Moerman17}, so that the Reynolds number ${\rey = \rho_o V R_0 / \eta_o}$ is small and inertia can essentially be neglected so that the velocity $\textbf{u}_{i,o}$ and pressure $P_{i,o}$, both inside and outside the droplet, satisfy Stokes' equations
\begin{equation}
  \label{eqs_flow}
  \nabla \cdot \textbf{u}_{i,o}= 0, \qquad 
  \nabla P_{i,o} = \eta_{i,o}\nabla^2 \textbf{u}_{i,o}.
\end{equation}
with subscripts $i,o$ referring to the inner and outer fluids, respectively. In the reference frame of the droplet's center considered here, the flow at infinity is opposite to the droplet's translation
\begin{equation}
  \label{far}
  \mathbf{u} \rightarrow -\mathcal{U}_\infty \mathbf{e}_z
  \qquad \textrm{for  } r \rightarrow \infty,
\end{equation}
and the system of equations above is closed by enforcing the mechanical equilibrium of the droplet in the absence of inertia, i.e. the force-free condition
\begin{equation}
  \label{forcefree}
  \int_{r=R}\boldsymbol\sigma_o\cdot\mathbf{n}\,\mathrm{d} S=\mathbf{0}.
\end{equation}

\subsection{Axisymmetric Stokes flow}
Axisymmetric Stokes flows inside and outside the droplet can be recast in terms of a streamfunction $\psi_{i,o}(t,r,\mu)$, such that
\begin{equation}
  \mathbf{u} = -\frac{1}{r^2} \frac{\partial \psi}{\partial \mu} \mathbf{e}_r
  - \frac{1}{r\sqrt{1-\mu^2}} \frac{\partial \psi}{\partial r} \mathbf{e}_\theta.
\end{equation} 
The general solution of the Stokes equations~(\ref{eqs_flow}) for the inner and outer flows in axisymmetric spherical coordinates is given by a superposition of orthogonal modes, the so-called Lamb solution~\cite{Lamb45, Happel83, Leal07}. The flow outside the droplet must converge to a finite unidirectional flow as $r \rightarrow \infty$, and the flow inside the droplet must be regular at the origin, so the streamfunction and pressure can be written generally in the outer fluid as
\begin{align}
  \label{outerflow}
  \psi_o \left( t, r, \mu \right)
  &= \left( \frac{a_{o,1}(t)}{r} - b_{o,1}(t) r^2 \right)
      \left( 1 - \mu^2 \right) 
  + \sum\limits_{n=2}^\infty 
      \left(\frac{a_{o,n}(t)}{r^n}-\frac{b_{o,n}(t)}{r^{n-2}}\right)
        \left( 1 - \mu^2 \right) L_n'(\mu), \\
  \label{outerpressure}
  P_o \left( t, r, \mu \right) 
  &= \eta_o \sum\limits_{n=2}^\infty 
    2 n \left( 1 - 2 n \right) b_{o,n}(t) \frac{L_n(\mu)}{r^{n+1}},
\end{align}
and within the droplet as
\begin{align}
  \label{innerflow}
  \psi_i \left( t, r, \mu \right)
  = &\sum\limits_{n=1}^\infty 
      \left( a_{i,n}(t) r^{n+1} - b_{i,n}(t) r^{n+3} \right)
        \left( 1 - \mu^2 \right) L_n'(\mu), \\
  \label{innerpressure}
  P_i \left( t, r, \mu \right) 
  = &- 2\eta_i \sum\limits_{n=0}^\infty 
     b_{i,n}(t) \left( n + 1 \right) 
      \left( 2 n + 3 \right) r^n L_n(\mu),
\end{align}
where $L_n$ denotes the $n$-th Legendre polynomial and prime denotes the derivative. Note that the Stokeslet term is omitted in~\eqref{outerflow} since the droplet is force-free~\cite{Blake70}, which effectively enforces~\eqref{forcefree} automatically. The different modes in~\eqref{outerflow} correspond to singularities of increasing order in the hydrodynamic signature of the swimming droplet, and are associated to specific physical characteristics of the self-propulsion and associated fluid motion. For instance, the mode with $n=1$ carries information about the droplet self-propulsion velocity (since it is the only mode with nonzero velocity as ${r \rightarrow \infty}$), mode with $n=2$ corresponds to a symmetric extensile flow akin to the flow excited by a force dipole (i.e. a stresslet), and so on. Note that the inner and outer hydrodynamic problems are fully determined by computing the intensity of the different modes $(a_{n,i},b_{n,i})$ and $(a_{n,o},b_{n,o})$, respectively.

\subsection{Nondimensionalization}
In the following, all quantities are non-dimensionalized using $R_0$, ${\cal A} R_0/{\cal D}$ and $R_0 / {\cal V}$ as reference scales for length, relative concentration of surfactant (i.e., $C-C_\infty$) and time, respectively. Here, ${\cal V}$ is the typical Marangoni velocity of a droplet in a surfactant gradient ${\cal A}$~\cite{Anderson89}
\begin{equation}
  \label{uInf_grad}
  {{\cal V} \equiv \frac{\gamma_C {\cal A} R_0}{{\cal D}(2 \eta_o + 3 \eta_i )}}.
\end{equation}
The pressure and viscous stress tensors are further rescaled as
\begin{equation}  
  P_i \rightarrow P_\infty + \dfrac{2 \gamma_0}{R_0} 
    + \frac{\eta_o {\cal V}}{R_0} \eta P_i,\quad
  P_o \rightarrow P_\infty + \frac{\eta_o {\cal V}}{R_0} P_o, \quad( \boldsymbol\tau_i, \boldsymbol\tau_o ) 
    \rightarrow \frac{\eta_o {\cal V}}{R_0} ( \eta \boldsymbol\tau_i, \boldsymbol\tau_o ),
\end{equation}
where $P_\infty$ is the constant background pressure, $\gamma_0$ the surface tension of the same spherical droplet in the absence of flow, and $\eta \equiv \eta_i / \eta_o$ is the viscosity ratio. We further note $U_\infty \equiv {\cal U}_\infty / {\cal V}$ the swimming velocity of the droplet. Besides $\eta$, the physical problem is entirely characterised by two additional non-dimensional parameters, the P\'eclet and capillary numbers, 
\begin{equation}
  \pe \equiv \frac{ {\cal V} R_0 }{ {\cal D} }, \qquad 
  \ca \equiv \frac{ \gamma_C {\cal A}R_0 }{ {\cal D} \gamma_0 }
\end{equation}
that characterize the relative magnitude of surfactant advection and diffusion, and the relative magnitude of Marangoni and Laplace stresses, respectively.

\subsection{Isotropic motionless base state}
Equations~\eqref{shape}--\eqref{forcefree} feature a motionless isotropic steady state given by,
\begin{equation}
\label{base}
  \bar{\xi} = 0, \; \;
  \bar{\textbf{u}}_i = \bar{\textbf{u}}_o = \textbf{0}, \; \;
  \bar{P_i} = \bar{P_o} = 0, \; \;
  \bar{C} = -1/r.
\end{equation}
Note that this solution~\eqref{base} exists for any values of $\pe$, $\ca$, and $\eta$; it features isotropic surfactant distribution, no Marangoni stresses, and, therefore, no droplet motion. In the following, we are interested in the existence and stability of additional non-isotropic steady states emerging in vicinity of the base state~\eqref{base}. This amounts mathematically to finding the fundamental eigenmodes of the system. The next section focuses on finding these eigenmodes and their existence condition (i.e., the corresponding value of $\pe$ for given $\ca$ and $\eta$), while in \S~\ref{weakly_nonlin} we investigate the steady flows sustained by nonlinear saturation of the eigenmodes. Finally, \S~\ref{lin1} analyses the stability of the trivial state~(\ref{base}) -- the stability of the non-isotropic steady state is presented in Appendix~\ref{lin2}.

\section{Neutrally stable eigenmodes of the linearized problem}
\label{neutral_modes}
We now carry out linear analysis of the problem stated in \S~\ref{statement}. Specifically, the dimensionless form of the problem formulated in~\eqref{shape}-\eqref{forcefree} are linearized about the base state~\eqref{base} in the case of a steady flow (i.e., $ \partial / \partial t = 0$). Solution of the resulting linear problem constitutes linear stability analysis of the base state in the limit of vanishing perturbation growth rates. By definition, perturbation growth rates vanish at the threshold of monotonic instability. Therefore, solution of the linearized problem allows us to (a) identify the instability threshold and (b) obtain the set of neutrally-stable eigenmodes. At the next stage of analysis, these eigenmodes are used to construct the steady flows emerging above the instability threshold (\S~\ref{weakly_nonlin}).

\subsection{Linearized equations}
The linearized advection-diffusion equation reads,
\begin{equation}
  \label{lin_ad}
   \nabla^2 C^{(1)}=-\frac{\pe}{r^4} \frac{\partial \psi_o^{(1)}}{\partial \mu},
\end{equation}
and linearized boundary conditions at the droplet interface write using domain perturbation, i.e., ${\left. f \right|_{r=R} \approx \left. f \right|_{r=1} + \xi \left. f' \right|_{r=1}}$,
\begin{align}
  \label{lin_bcs1}
  & \frac{\partial C^{(1)}}{\partial r} - 2 \xi^{(1)} = 0, \qquad
  \psi_i^{(1)} =\psi_o^{(1)} = 0, \qquad
  \frac{\partial \psi_i^{(1)}}{\partial r} 
    = \frac{\partial \psi_o^{(1)}}{\partial r}, \\
  \label{lin_bcs2}
  & \left( \frac{\partial^2}{\partial r^2} - 2 \frac{\partial}{\partial r}
    - \left( 1 - \mu^2 \right) \frac{\partial^2}{\partial \mu^2} \right) 
        \left( \psi_o^{(1)} - \eta \psi_i^{(1)} \right)
  = \left( 2 + 3 \eta \right) \left( 1 - \mu^2 \right) 
      \frac{\partial}{\partial \mu} \left( C^{(1)} + \xi^{(1)} \right), \\
  \label{lin_bcs3}
  & \frac{\ca}{2 + 3\eta} \bigg[
      - \eta P_i^{(1)} + P_o^{(1)}
      + 2 \left( \frac{\partial}{\partial r \partial \mu} 
        - 2 \frac{\partial}{\partial \mu} \right)
            \left( \psi_o^{(1)} - \eta \psi_i^{(1)} \right)
      \nonumber \\ &
      \qquad \qquad \qquad
      - 2 \left( 2 + 3 \eta \right) \left( C^{(1)} - 2 \xi^{(1)} \right)
    \bigg]
  = 2 \xi^{(1)} - 2 \mu \frac{\partial \xi^{(1)}}{\partial \mu}
  + \left( 1 - \mu^2 \right) \frac{\partial^2 \xi^{(1)}}{\partial \mu^2}.
\end{align}
where superscript $(1)$ denotes small perturbations of the base state. Because of the linearity of Stokes' equations, $\psi_{i,o}^{(1)}$ and $P_{i,o}^{(1)}$ assume the same form as~\eqref{outerflow}--\eqref{innerpressure}.

The streamfunction and pressure fields can be decomposed into orthogonal modes~\eqref{outerflow}--\eqref{innerpressure}. The form of the linearized equations~\eqref{lin_ad}--\eqref{lin_bcs3} suggests that the linearized concentration $C^{(1)}( r, \mu )$ and displacement $\xi^{(1)}( \mu )$ also decompose in orthogonal modes of the form
\begin{equation}
  \label{c1_xi1_modes_raw}
  C^{(1)}( r, \mu ) = \sum\limits_{n=0}^\infty C_n^{(1)} (r) L_n(\mu), \qquad
  \xi^{(1)}( \mu ) = \xi_0^{(1)} + \sum\limits_{n=2}^\infty \xi_n^{(1)} L_n(\mu),
\end{equation}
where the radial part of the basis functions of the concentration field $C_n^{(1)}( r )$ and constant amplitudes $\xi_n^{(1)}$, together with the coefficients ($a_{n,i}^{(1)} ,b_{n,o}^{(1)} ,b_{n,i}^{(1)} ,a_{n,o}^{(1)} $) are to be determined below and characterize each orthogonal eigenmode. In the expansion of $\xi^{(1)}( \mu )$ presented in~(\ref{c1_xi1_modes_raw}), we deliberately ignore the term $\propto L_1(\mu)$, since in the limit of small deformations this term corresponds to translation of the droplet, rather than deformation.

\subsection{Asymptotic structure of the concentration field}
\begin{figure}
\centering
  \includegraphics[scale=0.7]{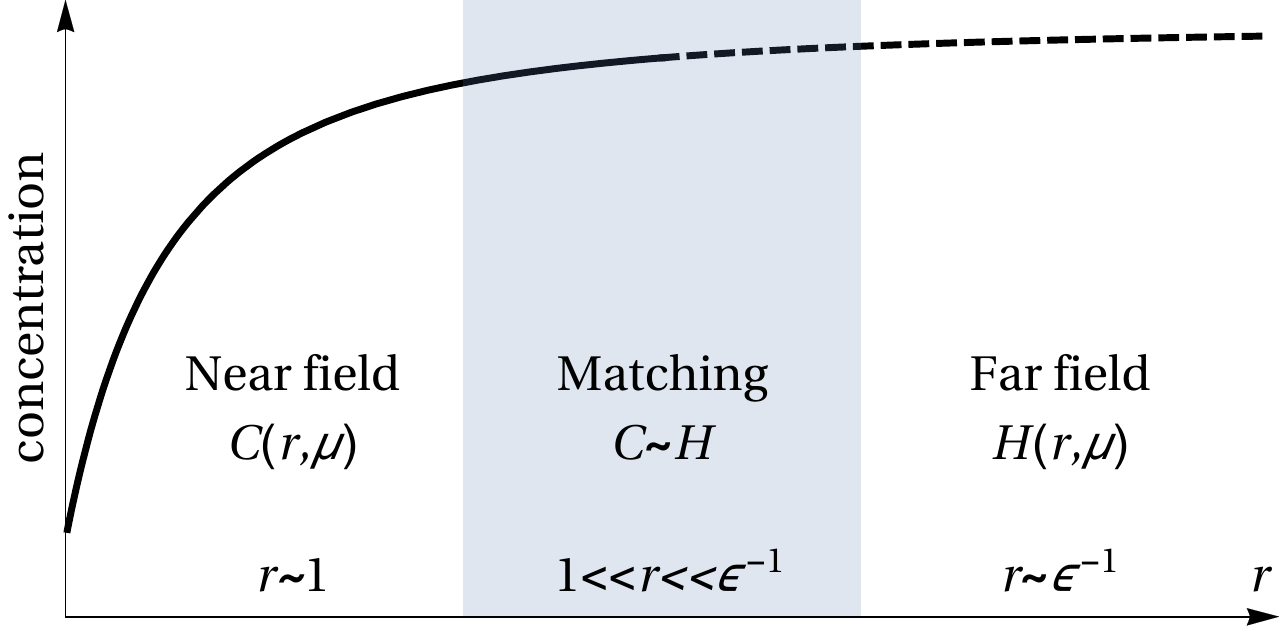}
  \caption{
    Asymptotic structure of the surfactant concentration field around the droplet. Near field solution $C(r,\mu)$ is matched with the far field solution $H(r,\mu)$ in the intermediate region where $1 \ll r \ll \epsilon^{-1}$.
  }
  \label{c_struct}
\end{figure}

Substitution of~\eqref{outerflow} and~\eqref{c1_xi1_modes_raw} into~\eqref{lin_ad} yields an equation for $C_n^{(1)}(r)$ admitting the following solution,
\begin{align}
  \label{c1_gen_sol1}
  & C_0^{(1)}(r) = \frac{c_0^{(1)}}{r} + d_0^{(1)}, \qquad
  C_1^{(1)}(r) = \frac{c_1^{(1)}}{r^2} + d_1^{(1)} r 
    + \pe \frac{a_{o,1}^{(1)} + 2 b_{o,1}^{(1)} r^3}{2 r^3}, \\
  \label{c1_gen_sol2}
  & C_n^{(1)}(r) \big|_{n > 1} = \frac{c_n^{(1)}}{r^{n+1}} + d_n^{(1)} r^n
    + \pe \frac{n a_{o,n}^{(1)} + (n + 1) b_{o,n}^{(1)} r^2}{2 r^{n+2}},
\end{align}
where $c_n^{(1)}$ and $d_n^{(1)}$ are unknown constants. In the case of a self-propelling drop (i.e. $b^{(1)}_{o,1}\neq 0$), this solution explicitly violates the far-field boundary condition~\eqref{far}. This is a well-known feature of advection-diffusion problems in the presence of a weak advective far-field flow: Acrivos and Taylor used matching asymptotic expansions to demonstrate that perturbations of the solute concentration field due to particle motion are dissipated in a boundary layer located at $r \sim 1/\epsilon\gg 1$, where $0 < \epsilon \ll 1$ quantifies the velocity of the particle with respect to the surrounding fluid~\cite{Acrivos62}. This framework applies here, since asymptotically small perturbations of a motionless base state are considered, and we thus aim to construct a composite solution for the concentration field consisting of two parts: (i) a near field part, corresponding to the immediate surroundings of the drop, $r \sim 1$,
\begin{equation}
  \label{c_near_exp}
  C(r,\mu) = -1/r + \epsilon C^{(1)} + \epsilon^2 C^{(2)} + \ldots,
\end{equation}
and (ii) a far field part valid away from the drop, $\rho\equiv r/\epsilon\sim 1$ ($r \gg 1$),
\begin{equation}
  \label{c_far_exp}
  H(\rho,\mu) = \epsilon H^{(1)} + \epsilon^2 H^{(2)} + \ldots,
\end{equation}
as shown in figure~\ref{c_struct}. Here $H(\rho,\mu)$ satisfies the rescaled advection-diffusion equation, namely,
\begin{equation}
  \label{c_far_eq}
  - \epsilon \pe \left(
        \frac{\partial \psi_o}{\partial \mu} \frac{\partial H}{\partial \rho}
      - \frac{\partial \psi_o}{\partial \rho} \frac{\partial H}{\partial \mu}
    \right)
  = \frac{\partial}{\partial \rho} 
      \left( \rho^2 \frac{\partial H}{\partial \rho} \right) 
  + \frac{\partial}{\partial \mu} 
      \left( \left( 1 - \mu^2 \right) \frac{\partial H}{\partial \mu} \right).
\end{equation}
Here $\epsilon$ is a small parameter that quantifies the distance to the isotropic steady state. Similarly to the work of Acrivos and Taylor~\cite{Acrivos62}, $C(r,\mu)$ encapsulates the dynamics of the droplet interface, whereas $H(\rho,\mu)$ is determined by advection of the concentration disturbances imparted by the droplet. Naturally, $C(r,\mu)$ and $H(\rho,\mu)$ must yield identical results in the matching region occurring at $1 \ll r \ll 1 / \epsilon$ (or, equivalently, $\epsilon \ll \rho \ll 1$)~\cite{Holmes95}. Note that although~\eqref{c_near_exp}--\eqref{c_far_exp} include higher-order terms in $\epsilon$, only $C^{(1)}$ and $H^{(1)}$ are relevant in the context of linear analysis.

\subsection{Concentration field far from the translating drop and asymptotic matching}
Linearization of the rescaled advection-diffusion equation~(\ref{c_far_eq}) about the base state, namely, about $\bar{H} = 0$ yields,
\begin{equation}
  \label{h1_ad_eq}
  F \left( H^{(1)} \right)  
  \equiv
  2 \pe \, b_{o,1}^{(1)} \left( \mu \frac{\partial H^{(1)}}{\partial \rho} 
    + \frac{1 - \mu^2}{\rho} \frac{\partial H^{(1)}}{\partial \mu} \right)
    + \frac{1}{\rho^2} \left[ 
        \frac{\partial}{\partial \rho} 
          \left( \rho^2 \frac{\partial H}{\partial \rho} \right) 
      + \frac{\partial}{\partial \mu} \left( \left( 1 - \mu^2 \right) 
          \frac{\partial H}{\partial \mu} \right)
    \right]
  = 0.
\end{equation}
Its solution that decays as $\rho \rightarrow \infty$ writes~\cite{Acrivos62}
\begin{equation}
  \label{h1_gen_sol}
  H^{(1)}( \rho, \mu ) = \frac{ e^{-\rho_s \mu} }{ \sqrt{ \left| \rho_s \right| } }
    \sum\limits_{n=0}^\infty h_n^{(1)} 
      K_{n+1/2} \left( \left| \rho_s \right| \right) L_n(\mu),
\end{equation}
where $h_n^{(1)}$ are unknown constants to be determined in the matching process with the inner solution, ${\rho_s \equiv \pe \, b_{o,1}^{(1)} \rho}$, and $K_n(x)$ denotes the modified Bessel function of the second kind of order $n$. We note that the direction of the droplet motion along the symmetry axis is determined by the sign of the constant amplitude $b_{o,1}^{(1)}$. Since there is no physical difference between the two directions of motion, we assume $b_{o,1}^{(1)} \geq 0$ in what follows.

We use Van Dyke's matching rule~\cite{Holmes95}, to match solutions~\eqref{c1_gen_sol1}--\eqref{c1_gen_sol2} and~(\ref{h1_gen_sol}) in the region ${\epsilon \ll \rho \ll 1}$. More specifically, $C( r, \mu )$ is expressed in terms of $\rho$ and both $C$ and $H$ are expanded in powers of $\rho\ll 1$. Note that only the terms linear in $\epsilon$, $\rho$, or $ \epsilon / \rho$ can be matched at the leading order of expansion. As a consequence,
\begin{equation}
\label{o1_match_cond}
  d_0^{(1)} = \pe \, b_{o,1}^{(1)}, \qquad
  h_0^{(1)} = -\sqrt{\frac{2}{\pi}} \pe \, b_{o,1}^{(1)}, \qquad
  d_n^{(1)} = h_n^{(1)} = 0 \qquad \text{for  } n > 0.
\end{equation}

\subsection{Solvability condition}
Substitution of $\psi_i^{(1)}$, $\psi_o^{(1)}$, and $C^{(1)}$ given by~\eqref{outerflow}, \eqref{innerflow} and~\eqref{c1_gen_sol1}--\eqref{c1_gen_sol2} into the boundary conditions~\eqref{lin_bcs1}--\eqref{lin_bcs3} and subsequent projection of the result onto the $n$-th Legendre polynomial yields a sequence of sets of homogeneous linear algebraic equations for the amplitudes $a_{i,n}^{(1)}$, $b_{i,n}^{(1)}$, $a_{o,n}^{(1)}$, $b_{o,n}^{(1)}$, $c_n^{(1)}$, and $\xi_n^{(1)}$. For given $n$, the solvability condition, i.e. the existence condition for a non-trivial solution to the linearized problem, reads
\begin{equation}
\label{pe_crit_def}
  \pe = \pe_n \equiv \begin{cases}
    4, & n = 1 \\
    \dfrac{4}{2 + 3 \eta} \left[
      \left( n + 1 \right)\left( 2 n + 1 \right)\left( 1 + \eta \right)
        - \ca \left( 1 + \eta \dfrac{n-1}{n+2} \right) 
      \right], & n > 1
  \end{cases}.
\end{equation}

\begin{figure}
\centering
  \raisebox{1.75in}{(a)}
  \includegraphics[scale=0.91]{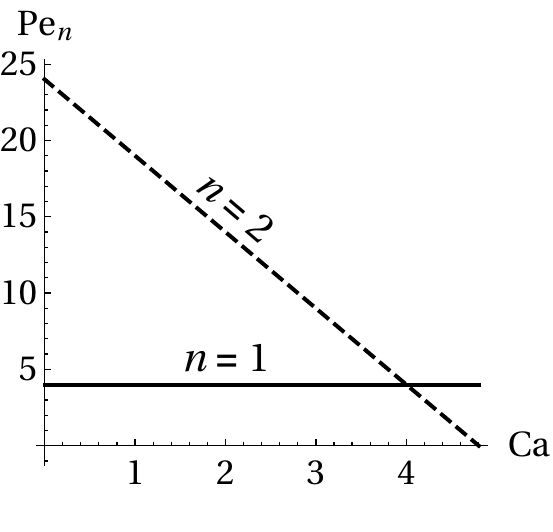}
  \quad
  \raisebox{1.75in}{(b)}
  \includegraphics[scale=0.55]{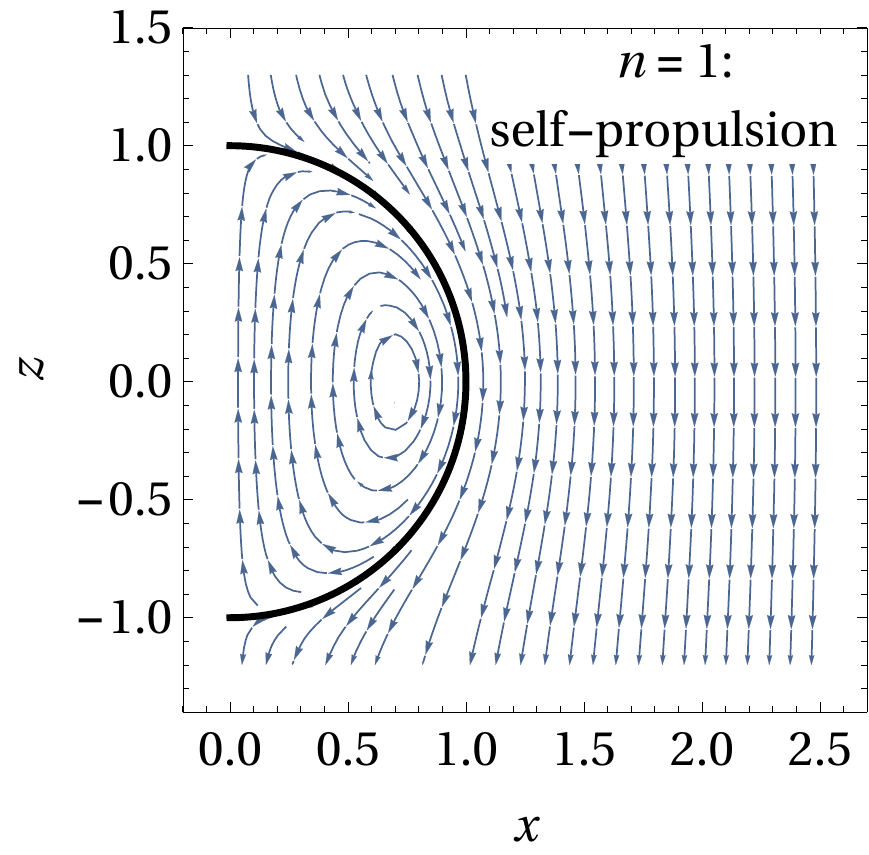}
  \quad
  \includegraphics[scale=0.55]{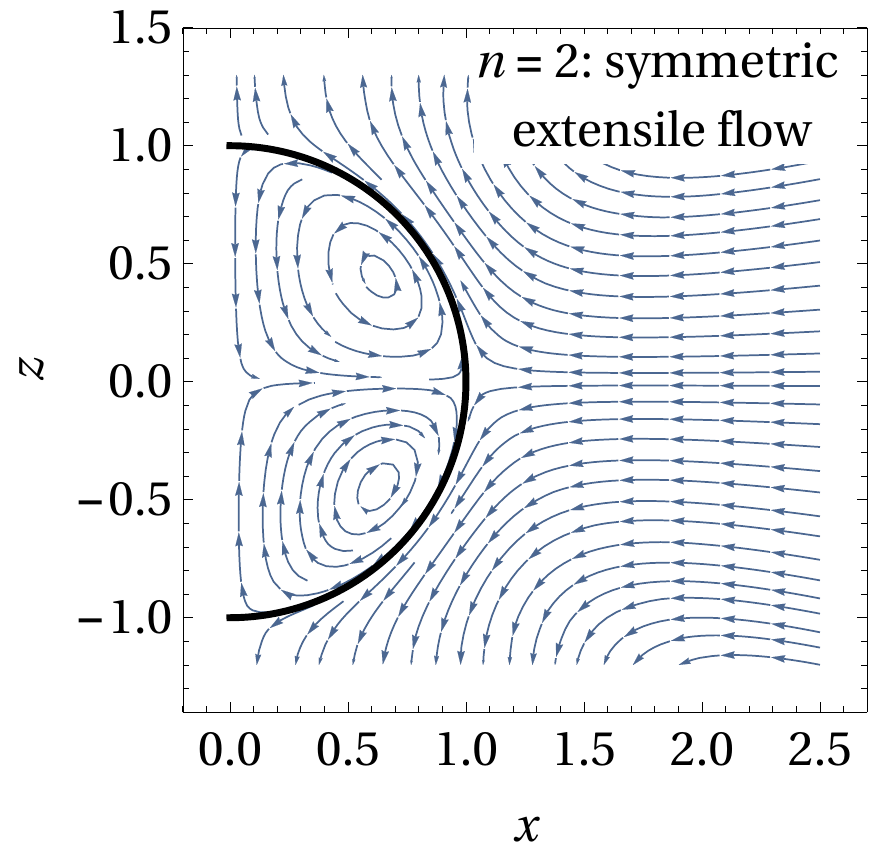}
  \caption{
    Two instability modes competing for the minimal value of the instability
    threshold $\pe_n$.
    (a) Evolution of the critical P{\'e}clet number with the capillary number 
      for $\eta = 1$.
    (b) Flow field corresponding to the first and the second mode of 
      instability, respectively.
  }
  \label{modes}
\end{figure}

Equations~(\ref{pe_crit_def}) determine when the linearized problem yields a non-trivial time-independent (i.e., neutrally stable) solution. In \S~\ref{lin1}, we will show that $\pe_n$ is the minimum P\'eclet number below (resp. above) which the isotropic state is linearly stable (resp. unstable) to perturbations along the $n$-th mode presented above. In other words, $\pe_n$ represents the threshold of the $n$-th mode of monotonic instability. This motivates referring to $\pe_n$ as the instability threshold for mode $n$ in the following. Depending on the value of the capillary number $\ca$, one of the first two modes in~\eqref{pe_crit_def} features the lowest instability threshold. As a consequence, the next stages of the analysis are focused exclusively on these first two instability modes, namely, the cases of $n=1$ and $n=2$, shown in figure~\ref{modes}.

The first eigenmode ($n=1$, $\pe_1=4$) reads
\begin{align}
  \label{o1_sol_n11}
  & b_{i,0}^{(1)} = \frac{\pe_1 A_1}{3 \eta} ( 2 + 3 \eta ), \qquad
  \frac{3}{2} a_{i,1}^{(1)} = \frac{3}{2} b_{i,1}^{(1)}
    = a_{o,1}^{(1)} = b_{o,1}^{(1)} = A_1, \\
  \label{o1_sol_n12}
  & c_0^{(1)} = 0, \qquad
  d_0^{(1)} = \pe_1 A_1, \qquad
  c_1^{(1)} = -\frac{3 \pe_1 A_1}{4}, \qquad
  d_1^{(1)} = 0,
\end{align}
and physically corresponds to a polar concentration field at the surface of the droplet $C^{(1)} \propto \mu$, which maintains its steady translation. The corresponding flow field is that of a translating droplet, i.e. the superposition of a steady flow with a source dipole singularity to enforce the impermeability condition. Note that the instability threshold $\pe_1=4$ does not include the viscosity ratio $\eta$, since we define dimensionless velocity based on the terminal velocity of a droplet in an imposed gradient of surfactant concentration~\eqref{uInf_grad}~\cite{Anderson89}.

The second eigenmode ($n=2$) can be written as
\begin{align}
  \label{o1_sol_n21}
  & a_{i,2}^{(1)} = b_{i,2}^{(1)} = a_{o,2}^{(1)} = b_{o,2}^{(1)} = A_2, \qquad
  \xi_2^{(1)} = \ca A_2 \frac{4 + \eta}{2 \left( 2 + 3 \eta \right)}, \\
  \label{o1_sol_n22}
  & c_2^{(1)} = -\frac{A_2}{3} \left( 7 \pe_2
    + \ca \frac{ 4 + \eta }{ 2 + 3 \eta } \right), \qquad
  d_2^{(1)} = 0.
\end{align}
In that case, the concentration field $C^{(1)}\propto L_2(\mu)$ is front-back symmetric and can not drive any net droplet motion. Instead, an extensile flow is forced by the Marangoni stress. Outside the droplet, it takes the same form as the second mode of the classical squirmer model~\cite{Blake70} and consists of a stresslet singularity (i.e., symmetric force dipole) and source quadrupole.

The amplitudes of the eigenmodes of the system, $A_1 \geq 0$ in~\eqref{o1_sol_n11}--\eqref{o1_sol_n12} or $A_2$ in~\eqref{o1_sol_n21}--\eqref{o1_sol_n22}, remain naturally undetermined within this linear framework. Weakly nonlinear analysis, which is the focus of the next section, will provide these saturation amplitudes in the vicinity of the critical conditions $\pe=\pe_n$.

Finally, we note that~\eqref{pe_crit_def} echoes the result of the stability analysis in the case of chemically active isotropic particles developed by Michelin~\textit{et al.}~\cite{Michelin13}. In particular, Michelin~{\it et al.} have demonstrated that the onset of spontaneous self-propulsion of active spherical particles also corresponds to $\pe = 4$. We argue that the persisting value of the instability threshold is related to the choice of dimensionless velocity. Both this paper and the work of Michelin~\textit{et al.}~\cite{Michelin13} define dimensionless velocity based on the velocity of an active drop/particle in an external concentration gradient $\cal A/\cal D$. In both cases, self-propulsion and the resulting advection of the concentration field generates a front-back concentration contrast. The onset of the drop or particle motion then corresponds in both cases to a fixed ratio of advective (i.e., destabilizing) and diffusive (i.e., stabilizing) terms, resulting in a fixed value of the P{\'e}clet number, $\pe = 4$.

\section{Weakly nonlinear analysis}
\label{weakly_nonlin}
Each of the neutrally stable eigenmodes obtained in \S~\ref{neutral_modes} exists at a distinct value of $\pe$ given by~\eqref{pe_crit_def}. To determine the saturation properties of the eigenmodes, we now successively analyse the behaviour of the different modes near the corresponding critical P{\'e}clet number. Formally speaking, higher-order terms of the asymptotic expansion established in~\eqref{c_near_exp}--\eqref{c_far_exp} are now included to investigate whether nonlinear terms allow for the saturation of growing perturbations, thus enabling a steady flow. Below, we focus specifically on the first two modes (which present the lowest critical P\'eclet number) and demonstrate that the two competing modes of instability shown in figure~\ref{modes} spawn two families of steady flows.

To study nonlinear behaviour of the neutral modes shown in figure~\ref{modes}, we assume that the P{\'e}clet number is close to the corresponding critical value,
\begin{equation}
  \label{pe_exp}
  \pe = \pe_n + \epsilon \delta,
\end{equation}
where $\epsilon \delta$ measures the distance to the critical P\'eclet number and $\delta = O\left( 1 \right)$. The concentration field is expanded as in~\eqref{c_near_exp}--\eqref{c_far_exp}, whereas the flow field and droplet shape are expanded near the isotropic steady state as
\begin{equation}
  \label{flow_xi_exp}
  \left( \psi_{i}, \psi_o, \xi \right)
  = \epsilon \left( 
      \psi_{i}^{(1)}, \psi_o^{(1)},  \xi^{(1)} \right)
  + \epsilon^2 \left( 
      \psi_{i}^{(2)}, \psi_o^{(2)},  \xi^{(2)} \right)
  + \ldots.
\end{equation}
At each order, approximation $\xi^{(j)}$ is given by a superposition of Legendre polynomials, as shown in~(\ref{c1_xi1_modes_raw}) and the streamfunctions are expanded as in~\eqref{outerflow} and \eqref{innerflow}. Substituting the expansions~\eqref{c_near_exp}--\eqref{c_far_exp}, \eqref{pe_exp}, and~\eqref{flow_xi_exp} into the dimensionless form of~\eqref{c_consump}--\eqref{forcefree}, a sequence of problems is obtained at successive orders of $\epsilon$. The first problem in the sequence comprises $O(\epsilon)$ terms and is identical to the linearized problem considered in \S~\ref{neutral_modes}. The rest of this section is devoted to the higher-order problems in $\epsilon$.

%
%
\subsection{Steady self-propulsion ($n=1$)}
\label{steady_n1_o2_o3}
In the case of $n=1$, ${\pe_1 \equiv 4}$ and the leading-order flow is given by the first squirming mode with coefficients presented in~\eqref{o1_sol_n11}--\eqref{o1_sol_n12}. This squirming mode corresponds to a self-propelling droplet and to determine the self-propulsion velocity, we now consider the problem at $\epsilon^2$ featuring quadratic interactions of the flow and concentration fields obtained at $\epsilon$.

\subsubsection{Concentration field around the droplet}
Quadratic approximation of the advection-diffusion equation (i.e., retaining only $O(\epsilon^2)$ terms) reads,
\begin{equation}
  \label{o2_ad}
  - \pe_n \frac{1}{r^4} \frac{\partial \psi_o^{(2)}}{\partial \mu} \
  - \nabla^2 C^{(2)}
  = \frac{\pe_n}{r^2} \left(
        \frac{\partial \psi_o^{(1)}}{\partial \mu} 
          \frac{\partial C^{(1)}}{\partial r}
      - \frac{\partial \psi_o^{(1)}}{\partial r}
          \frac{\partial C^{(1)}}{\partial \mu}
    \right)
  + \frac{\delta}{r^4} \frac{\partial \psi_o^{(1)}}{\partial \mu}.
\end{equation}
Using~\eqref{o1_sol_n11}--\eqref{o1_sol_n12}, and the resulting form of $C^{(1)}$ and $\psi_{i,o}^{(1)}$, the inhomogeneous right-hand side of~(\ref{o2_ad}) include nonzero projections onto the 0-th, 1-st, and 2-nd Legendre harmonics. Accordingly, the angular component of $C^{(2)}( r, \mu )$ is given by the first three Legendre polynomials, namely, $C^{(2)}( r, \mu ) = \sum\limits_{n=0}^2 C_n^{(2)}(r) L_n(\mu)$, with
\begin{align}
  \label{n1_c2_gen_sol1}
  & C_0^{(2)}(r) = \frac{c_0^{(2)}}{r} + d_0^{(2)} 
    - \pe_1^2 A_1^2 \frac{ 8 - 15 r + 20 r^3 + 80 r^6 }{120 r^5}, \\
  & C_1^{(2)}(r) = \frac{c_1^{(2)}}{r^2} + d_1^{(2)} r 
    + \delta A_1 \frac{ 1 + 2 r^3 }{2 r^3}
    + \pe_1 \frac{ a_{o,1}^{(2)} + 2 b_{o,1}^{(2)} r^3 }{2 r^3}, \\
  \label{n1_c2_gen_sol3}
  & C_2^{(2)}(r) = \frac{c_2^{(2)}}{r^3} + d_2^{(2)} r^2 
    + \pe_1 \frac{ 2 a_{o,2}^{(2)} + 3 b_{o,2}^{(2)} r^2 }{2 r^4}
    - \pe_1^2 A_1^2 \frac{ 10 - 21 r + 70 r^3 - 42 r^4 + 28 r^6 }{84 r^5},
\end{align}
where $c_n^{(2)}$ and $d_n^{(2)}$ are unknown constant amplitudes to be determined in the matching process with the far-field boundary layer.

Equation~\eqref{c_far_eq} reads at $O(\epsilon^2)$ as
\begin{equation}
  \label{n1_h2_ad_eq}
  F \left( H^{(2)} \right)
  = - 2 \left( \pe_1 b_{o,1}^{(2)} + \delta A_1 \right)
    \left( \mu \frac{\partial H^{(1)}}{\partial \rho} 
      + \frac{1 - \mu^2}{\rho} \frac{\partial H^{(1)}}{\partial \mu} \right).
\end{equation}
where the linear operator $F$ is defined in~\eqref{h1_ad_eq}. Since $A_1 \geq 0$, the solution of~\eqref{n1_h2_ad_eq} which decays as $\rho \rightarrow \infty$ reads,
\begin{equation}
  \label{n1_h2_gen_sol}
  H^{(2)}( \rho, \mu )
  = \frac{ \pe_1 b_{o,1}^{(2)} + \delta A_1 }{2}
    \left( 2 ( 1 + \mu ) + 3 \mu \frac{1 + \rho_s}{\rho_s^2} \right)
      e^{-\rho_s ( 1 + \mu ) } 
  + \frac{ e^{-\rho_s \mu} }{ \sqrt{\rho_s} } 
    \sum\limits_{n=0}^\infty h_n^{(2)} K_{n+1/2} \left( \rho_s \right) L_n(\mu),
\end{equation}
where $h_n^{(2)}$ are unknown constant amplitudes to be determined in the matching process, and ${\rho_s \equiv \pe_1 A_1 \rho \geq 0}$.

Asymptotic matching of $C(r,\mu)$ and $H(r,\mu)$ at $O(\epsilon^2)$ in the region $\epsilon \ll \rho \ll 1$ is achieved by expressing ${C( r, \mu ) = -1/r + \epsilon C^{(1)} + \epsilon^2 C^{(2)}}$ in terms of $\rho$, and expanding both $C( \rho, \mu )$ and $H( \rho, \mu )$ in powers of $\rho$. Linear and quadratic terms in $\epsilon$, $\rho$, or $\epsilon / \rho$ must now be matched, leading to
\begin{align}
\label{n1_o2_match_cond}
  & d_0^{(2)} = \pe_1 b_{o,1}^{(2)} + \delta A_1, \qquad
  d_1^{(2)} = -\pe_1^2 A_1^2, \qquad
  d_n^{(2)} = 0 \qquad \text{for  } n > 1, \\
  & h_1^{(2)} =  -3 \frac{ \pe_1 b_{o,1}^{(2)} + \delta A_1 }
    { \sqrt{ 2 \pi } }, \qquad
  h_n^{(2)} = 0 \qquad \text{for  } n \neq 1.
\end{align}

\subsubsection{Solvability condition}
Combination of~\eqref{n1_o2_match_cond} and~\eqref{n1_c2_gen_sol1}--\eqref{n1_c2_gen_sol3} yields $C^{(2)}( r, \mu )$. Expanding the boundary conditions~\eqref{c_consump} and~\eqref{stress_bc}--\eqref{kin_cont}, at $O(\epsilon^2)$ and projecting the result onto the first three Legendre polynomials, provides a set of inhomogeneous linear algebraic equations for the amplitudes $a_{i,n}^{(2)}$, $b_{i,n}^{(2)}$, $a_{o,n}^{(2)}$, $b_{o,n}^{(2)}$, $c_n^{(2)}$, and $\xi_n^{(2)}$. Solvability condition of this set of equations reads,
\begin{equation}
  \label{n1_o2_sol_cond}
  A_1 \left( \delta - 32 A_1 \right) = 0.
\end{equation}
Equation~(\ref{n1_o2_sol_cond}) implies that two branches of steady solutions of the nonlinear problem~\eqref{eqs_ad}--\eqref{forcefree}, exist near $\pe_1$: the first branch, given by $A_1 = 0$, corresponds to a motionless droplet and is in fact simply the isotropic steady state~\eqref{base} already discussed; the second branch describes a self-propelling drop with finite velocity (${A_1 = \delta / 32 > 0}$). Note that in the case of self-propulsion, the velocity of the droplet grows linearly with $\pe-\pe_1$ (i.e., $A_1 \propto \delta$) suggesting that the onset of the droplet motion is a transcritical bifurcation. Also recall that the self-propelling mode is associated with no droplet deformation. As a result, solvability condition~\eqref{n1_o2_sol_cond}, does not include $\ca$, i.e., in the leading order, deformability does not affect droplet self-propulsion velocity.

Following the steps of the analysis above, it is easy to demonstrate that in the case of $A_1 \leq 0$, the counterpart of the solvability condition~\eqref{n1_o2_sol_cond}, reads,
\begin{equation}
  \label{n1_o2_sol_cond_neg}
  A_1 \left( \delta + 32 A_1 \right) = 0,
\end{equation}
implying that, similarly to the case of $A_1 \geq 0$, the steady problem is solvable either for $A_1$ = 0 or for $\delta > 0$, i.e., above the threshold of Marangoni instability.

\subsubsection{Effect of deformability on the droplet's self-propulsion}
Equation~\eqref{n1_o2_sol_cond} provides the leading order evolution of the droplet velocity for the self-propelled steady state near the onset of propulsion, $\pe = \pe_1$, and was obtained by considering the solvability condition of the problem including corrections up to $O(\epsilon^2)$. We now use the same approach to extend the expansion of the different equations up to $O(\epsilon^3)$ and obtain the quadratic correction for self-propulsion velocity.

When the solvability condition~\eqref{n1_o2_sol_cond} is satisfied, the non-trivial solution of the problem at $O(\epsilon^2)$ writes,
\begin{align}
  \label{n1_o2_sol1}
  & b_{i,0}^{(2)} = \dfrac{\left( 2 + 3 \eta \right)
    \left( 29 \delta^2 + 10240 B_1 \right)}{7680 \eta}, \qquad
  \frac{3}{2} a_{i,1}^{(2)} = \frac{3}{2} b_{i,1}^{(2)} 
    = a_{o,1}^{(2)} = b_{o,1}^{(2)} = B_1, \\
  & a_{i,2}^{(2)} = b_{i,2}^{(2)} = a_{o,2}^{(2)} = b_{o,2}^{(2)}
    = -\frac{33 \delta^2}{896 \left( \pe_2 - 4 \right)}, \qquad
  \xi_2^{(2)} = -\frac{33 \ca \delta^2 ( 4 + \eta )}
    {1792 \left( 2 + 3 \eta \right) \left( \pe_2 - 4 \right)}, \\
  & c_0^{(2)} = -\frac{\delta^2}{128}, \qquad
  d_0^{(2)} = 4 B_1 + \frac{\delta^2}{32}, \qquad
  c_1^{(2)} = -\left( \frac{\delta^2}{32} + 3 B_1 \right), \qquad
  d_1^{(2)} = -\frac{\delta^2}{64}, \\
  \label{n1_o2_sol4}
  & c_2^{(2)} = 3 \delta^2 \frac{20 ( 12 + 17 \eta ) + 3 \ca ( 4 + \eta )}
    {896 \left( 2 + 3 \eta \right) \left( \pe_2 - 4 \right)}, \qquad
  d_2^{(2)} = 0,
\end{align}
where the remaining unknown constant $B_1$ is determined from the solvability condition at O$(\epsilon^3)$. It should be noted in~\eqref{n1_o2_sol1}--\eqref{n1_o2_sol4} that the leading order deformation of the self-propelling droplet is $O(\epsilon^2)$ (recall that $\xi^{(1)}=0$) and always corresponds to an oblate shape ($\xi_2^{(2)}<0$). The leading order volume change is $O(\epsilon^4)$, so that volume conservation is automatically enforced up to quartic order in $\epsilon$.

\begin{figure}
  \centering
  \includegraphics[scale=1.1]{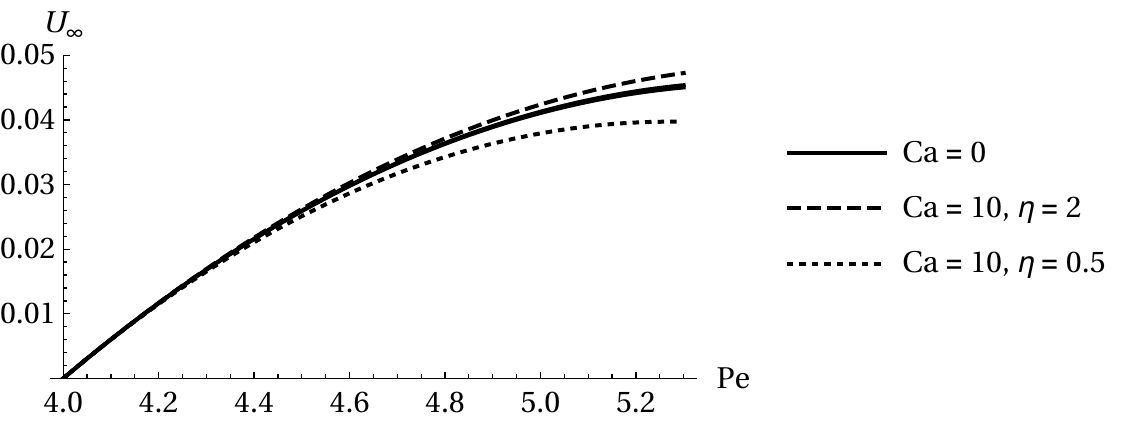}
  \caption{
    Effect of deformability ($\ca$) and relative viscosity ($\eta$) on the 
    self-propulsion velocity of an active droplet near the onset of Marangoni 
    instability as given in~(\ref{n1_o3_velocity}).
    The solid line represents the case of a nondeformable droplet ($\ca = 0$) 
    with $\eta = 0.5$ or $\eta = 2$ 
    (both cases are essentially indistinguishable due to weak dependence of 
    $U_\infty$ on $\eta$ in the limit of $\ca = 0$).
    Dotted and dashed lines represent deformable droplets ($\ca = 10$), 
    with $\eta = 0.5$ and $\eta = 2$, respectively.
  }
  \label{propulsion_u}
\end{figure}

To obtain a correction to the droplet self-propulsion velocity, the weakly non-linear analysis must be carried out up to the cubic order. The solution procedure of the problem at $\epsilon^3$ remains exactly the same as for the lower-order problems, and several intermediate results of the derivation are provided in Appendix~\ref{details}. As for the $\epsilon^2$-problem, solvability condition at $\epsilon^3$ provides information about the $O(\epsilon^2)$ droplet velocity, namely,
\begin{equation}
  \label{n1_o3_velocity}
  U_\infty 
  = 2 \left( \epsilon A_1 + \epsilon^2 B_1 \right)
  = \frac{\pe - 4}{16}
    - \frac{(\pe - 4)^2 
      \big[ 4 \left( 2 + 3 \eta \right) \left( 343073 + 325872 \eta \right)
        - 35 \ca \left( 4 + \eta \right) \left( 932 + 2883 \eta \right) \big]}
    {1254400 \left( 2 + 3 \eta \right)^2 \left( \pe_2 - 4 \right)},
\end{equation}
with ${\pe_2(\ca,\eta)=[60(1+\eta)-\ca(4+\eta)]/(2+3\eta)}$ (see~\eqref{pe_crit_def}). One can immediately observe that at this order ${\partial_\ca U_\infty > 0}$ (resp.~${\partial_\ca U_\infty < 0}$) when ${\eta > 1.0134}$ (resp.~${\eta < 1.0134}$). The deformability of the droplet's interface therefore affects differently the self-propulsion of droplets that are more or less viscous than the surrounding fluid: roughly speaking, equation~\eqref{n1_o3_velocity} states that deformability enhances self-propulsion of viscous droplets, but hinders propulsion of drops that are less viscous than the surrounding fluid (figure~\ref{propulsion_u}).

This result can be interpreted as follows. Steady self-propulsion of the droplet occurs when the Stokes drag balances the thrust generated by Marangoni stresses, that result from concentration gradients at the surface. In our analysis, this balance is represented by a saturated self-propelling eigenmode, where the saturation comes from nonlinear terms implementing weakly nonlinear interaction of the different components of the solution. At $\epsilon^3$, cumulative Marangoni stress on the drop is ${\propto C_1^{(3)} ( r = 1 )}$, see~\eqref{n1_c3_gen_sol2}, which includes the term $\propto 1 / \left( \pe_2 - 4 \right)$ resulting from the interaction of the front-back symmetric component ($n=2$) of the $O(\epsilon^2)$ concentration mode with the leading-order (i.e., $O(\epsilon)$) flow associated with self-propulsion. $\pe_2$ is a decreasing function of $\ca$, thus increasing deformability enhances the term $\propto 1 / \left( \pe_2 - 4 \right)$; moreover, when $r=1$, this term is positive (resp. negative) for $\eta < 39 / 29$ (resp. $\eta>39/29$). That is, increase in capillary number tend to increase the front-back concentration gradient of surfactant (and Marangoni forcing) for viscous droplets, while deformability tend to reduce them for less viscous droplets.

In addition, equations~\eqref{n1_o2_sol1}--\eqref{n1_o2_sol4} establish that self-propulsion at ${\ca > 0}$ is always accompanied by droplet deformations of order $\epsilon^2$ with an oblate shape, ${\xi^{(2)} < 0}$. Oblate deformations are known to increase the Stokes drag on a steadily moving droplet~\cite{Matunobu66}, namely,
\begin{equation}
  F_D \propto 1 - \epsilon^2 \xi_2^{(2)} \frac{3 \eta^2 - \eta + 8}
    {5 ( 1 + \eta ) ( 2 + 3 \eta )}.
\end{equation}
In our case, magnitude of droplet deformations given by $\left| \xi^{(2)} \right|$ decreases with increasing $\eta$, i.e., less viscous droplets deform more and experience higher drag, and a lower velocity for a given Marangoni forcing. Both effects of $\ca$ (modified Marangoni forcing and modified viscous drag) therefore reinforce each other: drops with smaller $\eta$ deform more than their viscous counterparts and, thus, experience higher drag, and they also experience a reduced Marangoni forcing, leading to the non-monotonous effect of deformability illustrated in figure~\ref{propulsion_u}. Interestingly, oblate deformations of active droplets were also predicted in the limit of ${\pe \gg 1}$ by Golovin~\textit{et al.}~\cite{Golovin89} and in the presence of a chemical reaction in the bulk fluid by Yoshinaga~\cite{Yoshinaga14}. An alternative approach based on reaction-diffusion equations also yields oblate deformations of the reacting domain~\cite{Shitara11}.

%
%
\subsection{Symmetric extensile flow ($n=2$)}
\label{steady_symm}
The previous section focused on the analysis of the self-propelling mode of instability ($n=1$). The present section focuses now on the second mode characterized by a symmetric extensile flow (see figure~\ref{modes}). In the case of $n = 2$, solvability condition~\eqref{pe_crit_def} yields, $\pe = \pe_2$, and the leading-order flow is given by the second squirming mode with an unknown amplitude $A_2$ introduced in~\eqref{o1_sol_n21}--\eqref{o1_sol_n22}. In order to determine $A_2$, the solvability condition must be established for the $\epsilon^2$-problem  featuring quadratic interactions of the flow and concentration fields obtained in \S~\ref{neutral_modes}.

\subsubsection{Concentration distribution around the droplet}
For $n=2$, ${C^{(1)}( r, \mu ) \propto L_2(\mu)}$, and quadratic nonlinearities in~(\ref{o2_ad}) produce nonzero projections of the $O(\epsilon^2)$ concentration field onto the 0-th, 2-nd, and 4-th Legendre harmonics, that is,  
\begin{equation}
  \label{n2_c2_exp}
  C^{(2)}( r, \mu ) = \sum\limits_{n=0}^2 C_{2n}^{(2)}(r) L_{2n}(\mu),
\end{equation}
with
\begin{align}
\label{n2_c2_gen_sol}
  C_0^{(2)}(r) = & \, \frac{c_0^{(2)}}{r} 
    - \pe_2^2 A_2^2 \frac{ 180 - 490 r + 126 r^2 + 735 r^3 - 630 r^4 }{1050 r^7}
    \nonumber \\ &
    \qquad \qquad \qquad \qquad \qquad \qquad \qquad \qquad
    + \pe_2 \ca A_2^2 \frac{\left( 4 + \eta \right) \left( 2 - 3 r^2 \right)}
        {30 r^6 \left( 2 + 3 \eta \right) }, \\
  C_2^{(2)}(r) = & \, \frac{c_2^{(2)}}{r^3} 
    + \delta A_2 \frac{2 + 3 r^2}{2 r^4}
    + \pe_2 \frac{ 2 a_{o,2}^{(2)} + 3 b_{o,2}^{(2)} r^2 }{2 r^4}
    + 2 \pe_2 \ca A_2^2 \frac{\left( 4 + \eta \right) \left( 1 - 3 r^2 \right)}
        {21 r^6 \left( 2 + 3 \eta \right)}
    \nonumber \\ & - \pe_2^2 A_2^2 \frac{ 875 - 2450 r + 225 r^2 + 7350 r^3
      + 756 r^4 + 3780 r^4 \log r }{3675 r^7}, \\
  C_4^{(2)}(r) = & \, \frac{c_4^{(2)}}{r^5} 
    + \pe_2 \frac{ 4 a_{o,4}^{(2)} + 5 b_{o,4}^{(2)} r^2 }{2 r^6}
    + 3 \pe_2 \ca A_2^2 \frac{\left( 4 + \eta \right) \left( 4 + 9 r^2 \right)}
        {70 r^6 \left( 2 + 3 \eta \right)}
    \nonumber \\ & - \pe_2^2 A_2^2 \frac{ 2016 - 6468 r + 616 r^2 
      - 14553 r^3 + 3564 r^4 + 5544 r^2 \log r }{5390 r^7},
\end{align}
where $c_n^{(2)}$ are constant amplitudes to be determined. In the absence of any translation of the droplet, rescaled advection-diffusion equation~(\ref{c_far_eq}) at $\epsilon^2$ reduces to Laplace's equation, implying that for $n=2$ advection of the perturbations in the far field is negligible (i.e., no boundary layer is needed here). Equivalently, it is now possible to satisfy the far-field and near field boundary conditions for $C^{(2)}(r,\mu)$, which  must decay as $r \rightarrow \infty$.

\subsubsection{Boundary and solvability conditions}
Expanding the boundary conditions~\eqref{c_consump} and~\eqref{stress_bc}--\eqref{kin_cont}, up to $O(\epsilon^2)$, and substituting for $C^{(1)}$, $C^{(2)}$, and the corresponding flow field, projection of the result onto the first five Legendre polynomials provides a set of inhomogeneous linear algebraic equations for the amplitudes $a_{i,n}^{(2)}$, $b_{i,n}^{(2)}$, $a_{o,n}^{(2)}$, $b_{o,n}^{(2)}$, $c_n^{(2)}$, and $\xi_n^{(2)}$. In particular, projection of the kinematic boundary condition onto $L_0(\mu)$ yields,
\begin{equation}
  \label{n2_o2_kin}
  \ca A_2^2 \frac{6 ( 4 + \eta )}{5 ( 2 + 3 \eta )} = 0.
\end{equation}
That is, symmetric steady flow field around an active droplet is not possible for $\ca=O(1)$, but may exist in the limit of a weakly deformable droplet, i.e., $\ca \sim O(\epsilon)$. Indeed, a weakly deformable droplet is spherical in the leading order, $\xi^{(1)} = 0$, and features instability thresholds independent of $\ca$. In the particular case of $\ca \sim \epsilon$, the $O(\ca)$ terms are pushed to the next order of asymptotic expansion. For instance, the $O(\ca)$ terms in~\eqref{lin_bcs1}--\eqref{lin_bcs3} appear in the corresponding boundary conditions of the problem at $\epsilon^2$. Consequently, the solvability condition of the $\epsilon^2$-problem is
\begin{equation}
  \label{n2_o2_sol_cond}
  A_2 \left( 49 \delta \left( 2 + 3 \eta \right)^2 
    + 49 \ca \left( 4 + \eta \right) \left( 2 + 3 \eta \right)
    + 16320 A_2 \left( 1 + \eta \right)^2 \right) = 0.
\end{equation}
Equation~(\ref{n2_o2_sol_cond}) establishes that in the limit of a weakly deformable active droplet, two branches of steady solutions exist in the $\epsilon$-neighborhood of $\pe_2$: the first branch, given by $A_2 = 0$, corresponds to a motionless state; whereas the second, featuring ${A_2 < 0}$, describes a symmetric extensional flow field akin to that of a force dipole. Recall that equation~\eqref{o1_sol_n22} connects the coefficient $A_2$ with a particular type of droplet deformation: ${A_2 > 0}$ (resp. ${A_2 < 0}$) corresponds to prolate (resp. oblate) deformations. In general, prolate and oblate deformations are not symmetric, so it is natural that our analysis yields $A_2$ with a particular sign.

%
%
\subsection{Simultaneous onset of the two dominant instability modes ($\pe_1 \sim \pe_2$)}
\label{steady_comp}
Definition of the instability thresholds~(\ref{pe_crit_def}) implies that for
\begin{equation}
  \label{def_cond}
  \ca = 4 ( 13 + 12 \eta ) / ( 4 + \eta ) + O(\epsilon)
\end{equation}
the thresholds of the first two eigenmodes coincide, namely, ${|\pe_1 - \pe_2| = O(\epsilon)}$. The purpose of this section is to investigate how the potential interaction of these two modes may impact the self-propulsion of the droplet. In this case, the leading order flow field is given by the first two squirming modes with coefficients presented in~\eqref{o1_sol_n11}--\eqref{o1_sol_n22}, respectively. We now consider the problem at $\epsilon^2$ featuring quadratic interactions of the two instability modes shown in figure~\ref{modes}.

Weakly nonlinear dynamics of the system is investigated in the case of 
\begin{equation}
  \ca = 4 ( 13 + 12 \eta ) / ( 4 + \eta ) + \epsilon \ca_1, 
\end{equation}
where the system admits two linearly-independent eigenmodes and $C^{(1)}( r, \mu )$ is represented by a combination of $L_0(\mu)$, $L_1(\mu)$, and $L_2(\mu)$. Thus, at $\epsilon^2$ quadratic nonlinearities in~(\ref{o2_ad}) produce nonzero projections onto the first five Legendre harmonics. Following the steps of the analysis presented in \S~\ref{steady_n1_o2_o3}--\ref{steady_symm}, the rescaled advection-diffusion equation~\eqref{c_far_eq} reduces at $\epsilon^2$ to~(\ref{n1_h2_ad_eq}) with a solution given by~\eqref{n1_h2_gen_sol}, and near and far field solutions match in the region $\epsilon \ll \rho \ll 1$, when conditions~\eqref{n1_o2_match_cond} are met.

Expanding the boundary conditions at the droplet interface to $O(\epsilon^2)$, the projection of the kinematic boundary condition onto $L_0(\mu)$ yields
\begin{equation}
  \label{n12_o2_kin}
  \frac{24 ( 13 + 12 \eta )}{5 ( 2 + 3 \eta )} A_2^2 = 0,
\end{equation}
and can only be satisfied when $A_2 = 0$, as for the case of mode $n=2$ alone, see~\eqref{n2_o2_kin}. Consequently, the solvability condition for the problem at $\epsilon^2$ in the case of $\pe_1 \sim \pe_2$ reads,
\begin{equation}
  \label{n12_o2_sol_cond}
  A_1 = 0.
\end{equation}
No nontrivial solution can be found within the $O(\epsilon)$ neighbourhood of $\pe_1\sim\pe_2$, and in particular the regime of steady self-propulsion discussed in \S~\ref{steady_n1_o2_o3} ceases to exist due to the competition between the first and the second modes of Marangoni instability. In other words, we have arrived to a conclusion that deformability may cause a qualitative change in droplet dynamics: droplets with ${\ca < 4 ( 13 + 12 \eta ) / ( 4 + \eta )}$ exhibit a regime of steady self-propulsion with ${U_\infty \propto \pe - \pe_1}$, whereas highly-deformable drops with ${\ca \geq 4 ( 13 + 12 \eta ) / ( 4 + \eta )}$ have no steady regime to reach in vicinity of the base state~\eqref{base}. As demonstrated in the following section, this qualitative change in droplet behaviour can be linked to the asymptotic disparity of the time scales associated with the first two modes of the Marangoni instability.

\section{Linear stability analysis and growth rates}
\label{lin1}
The analysis developed so far was focused on the steady flows emerging due to saturation of neutrally-stable modes. To gain further insight on the different transitions identified and the stability of each of these states (including the isotropic base state), the linear stability analysis of the system is now carried out in the vicinity of the steady solutions obtained in the previous section. We focus primarily on the linear stability of the isotropic base state~\eqref{base} the stability of the self-propelled mode being presented in Appendix~\ref{lin2}. Specifically, we obtain the growth rates of the system's eigenmodes identified in~\eqref{o1_sol_n11}--\eqref{o1_sol_n12} and~\eqref{o1_sol_n21}--\eqref{o1_sol_n22} and shown in figure~\ref{modes}. This analysis finally demonstrates that near their respective instability threshold, the second instability mode grows asymptotically faster than the first one. This disparity results in fact from these modes being associated with fundamentally-different physical phenomena: the first mode is intrinsically-linked to the symmetry breaking of the advective boundary layer far from the droplet, whereas the second mode depends only on the interfacial dynamics of the drop.

Equations~\eqref{shape}--\eqref{forcefree} are first linearized around the isotropic base state~\eqref{base} introducing the time-dependent normal perturbations
\begin{subequations}
\label{lin1_pert}
\begin{gather}
  \left( \psi_i, P_i, \psi_o, P_o \right)(t,r,\mu)
  = e^{\lambda t} \left( \tilde{\psi}_i, \tilde{P}_i,
        \tilde{\psi}_o, \tilde{P}_o \right)(r,\mu), \\
  C( t, r, \mu ) = -\frac{1}{r} + e^{\lambda t} \tilde{C} ( r, \mu ), \; \;
  H( t, \rho, \mu ) = - \frac{\epsilon}{\rho} + e^{\lambda t} \tilde{H} ( \rho, \mu ), \; \;
  \xi( t, \mu ) = e^{\lambda t} \tilde{\xi}(\mu),
\end{gather}
\end{subequations}
where tilde denotes the perturbations and $\lambda$ is the perturbations' growth rate. For simplicity, we only consider monotonically unstable case, namely, $\lambda > 0$. Linearization for fixed values of $\pe$, $\ca$, and $\eta$ produces a linear eigenvalue problem for $\lambda$ and the associated eigenmode ($\tilde{\psi}_i$, $\tilde{P}_i$, $\tilde{\psi}_o$, $\tilde{P}_o$, $\tilde{C}$, $\tilde{H}$, $\tilde{\xi}$). Note that this is in fact simply the generalization of~\eqref{lin_ad}--\eqref{lin_bcs3} to the time-dependent perturbations.

As for the steady linear analysis, orthogonal eigenmodes take the form in~\eqref{outerflow}, \eqref{innerflow} and~\eqref{c1_xi1_modes_raw}. In particular, projecting the linearized advection-diffusion equation along the first two Legendre polynomials leads to 
\begin{align}
  \label{lin1_ad1}
  & \frac{\partial}{\partial r} 
    \left( r^2 \frac{\partial \tilde{C}_1}{\partial r} \right)  
      - \left( \pe \lambda r^2 + 2 \right) \tilde{C}_1 
  = 2 \pe \frac{\tilde{a}_{o,1} - \tilde{b}_{o,1} r^3}{r^3}, \\
  \label{lin1_ad2}
  & \frac{\partial}{\partial r} 
    \left( r^2 \frac{\partial \tilde{C}_2}{\partial r}  \right)  
      - \left( \pe \lambda r^2 + 6 \right) \tilde{C}_2 
  = 6 \pe \frac{\tilde{a}_{o,2} - \tilde{b}_{o,2} r^2}{r^4},
\end{align}
where $\tilde{C}_n$ denotes the $n$-th eigenmode of $\tilde{C}$ and $\tilde{a}_{o,1}$, $\tilde{b}_{o,1}$, $\tilde{a}_{o,2}$, and $\tilde{b}_{o,2}$ denote the amplitudes of the first two squirming modes in the expansion of $\tilde{\psi}_o$.

Unlike their steady counterparts, equations~\eqref{lin1_ad1}--\eqref{lin1_ad2}, which govern the radial component of $\tilde{C}_1$ and $\tilde{C}_2$,  allow for an exponential decay of the surfactant concentration as $r \rightarrow \infty$. Therefore, there is no need to consider a far field solution separately. Combined with the far field boundary condition~\eqref{lin1_ad1}--\eqref{lin1_ad2} yield the following expressions for the first two eigenmodes of $\tilde{C}$,
\begin{align}
  \label{lin1_c1_sol}
  & \tilde{C}_1 
  = \tilde{c}_1 e^{-\lambda_s r} \frac{1 + \lambda_s r}{r^2}
  + \frac{\pe \tilde{a}_{o,1}}{2 r^3} + \frac{2 \tilde{b}_{o,1}}{\lambda r^2}
  \nonumber \\ & \qquad \qquad \qquad
  - \frac{\pe \lambda_s \tilde{a}_{o,1}}{8 r^2} \left[ 
      \left( 1 + \lambda_s r \right) e^{-\lambda_s r} 
        \ei \left( \lambda_s r \right)
      - \left( 1 - \lambda_s r \right) e^{\lambda_s r} 
          \ei \left( -\lambda_s r \right)
    \right], \\
  \label{lin1_c2_sol}
  & \tilde{C}_2
  = \tilde{c}_2 e^{-\lambda_s r} 
      \frac{3 + 3 \lambda_s r + \lambda_s^2 r^2}{r^3}
  + \pe \tilde{a}_{o,2} \frac{8 + \lambda_s^2 r^2}{8 r^4}
  - \frac{3 \pe \tilde{b}_{o,2}}{4 r^2}
  + \pe \frac{6 \tilde{b}_{o,2} - \lambda_s^2 \tilde{a}_{o,2}}{16 \lambda_s r^3} 
    \nonumber \\ & \qquad \quad \times
    \left[ 
      \left( 3 + 3 \lambda_s r + \lambda_s^2 r^2 \right) e^{-\lambda_s r} 
        \ei \left( \lambda_s r \right)
      - \left( 3 - 3 \lambda_s r + \lambda_s^2 r^2 \right) e^{\lambda_s r} 
          \ei \left( -\lambda_s r \right)
    \right],
\end{align}
where $\lambda_s \equiv \sqrt{ \pe \lambda }$, $\ei(x)$ denotes the exponential integral, and constants $\tilde{c}_n$ are to be determined from the boundary conditions at the droplet interface.

Substituting the eigenmodes of $\tilde{\psi}_i$, $\tilde{P}_i$, $\tilde{\psi}_o$, and $\tilde{P}_o$ along with~\eqref{lin1_c1_sol}--\eqref{lin1_c2_sol} into the linearized boundary conditions~\eqref{c_consump} and~\eqref{stress_bc}--\eqref{kin_cont}, two sets of linear algebraic equations are obtained. In turn, solvability conditions of these sets determine the respective growth rate of the corresponding perturbation. In the case of the first instability mode, solvability condition is identical to equation~(14) from Ref.~\cite{Michelin13} and can be simplified for ${|\pe - \pe_1| \ll \pe_1}$ (recall that ${\pe_1 = 4}$, see~\eqref{pe_crit_def}), resulting in the following leading order behaviour for the growth rate of the first mode:
\begin{equation}
  \label{lin1_n1_sol_cond}
  \lambda_{s,1} = \frac{3 ( \pe - \pe_1 )}{16}.
\end{equation}
Recall that we assumed $\lambda > 0$, that is, solvability condition~(\ref{lin1_n1_sol_cond}) holds only for $\pe > \pe_1$, where the motionless steady state becomes unstable with respect to the first mode of instability.

Solvability condition of the second instability mode is also treated asymptotically for ${| \pe - \pe_2 | \ll \pe_2}$, with $\pe_2$ defined in~\eqref{pe_crit_def}. A different leading order scaling is obtained this time, namely ${\lambda_{s,2} \propto \sqrt{ \pe - \pe_2 }}$, and the leading order growth rate for the second mode near $\pe=\pe_2$ finally reads,
\begin{equation}
  \label{lin1_n2_sol_cond}
  \lambda_{s,2} = \sqrt{ 
    \frac{ 24 \pe_2 ( 2 + 3 \eta ) ( \pe - \pe_2 ) }
      {38400 \left( 1 + \eta \right)^2
        - 4 \ca ( 4 + \eta ) ( 292 + 283 \eta ) 
          + \ca^2 ( 4 + \eta ) ( 64 + 31 \eta )} }.
\end{equation}
Equation~\eqref{lin1_n2_sol_cond} holds only for $\pe > \pe_2$, when the motionless base state becomes unstable with respect to the second mode of instability.

Equations~(\ref{lin1_n1_sol_cond}) and~(\ref{lin1_n2_sol_cond}) imply that near their respective thresholds, the second mode of instability grows asymptotically faster than the first one,
\begin{equation}
  \label{lin1_n1_n2}
  \lambda_{s,2} \gg \lambda_{s,1}.
\end{equation}
This result is particularly important in the case of $\pe_1 \sim \pe_2$, when the first two instability modes are excited simultaneously. In particular, equation~\eqref{lin1_n1_n2} suggests that for $\pe_1 \sim \pe_2$, saturation of the self-propelling mode is asymptotically slower, compared to the excitation of the symmetric extensile flow associated with the second mode. We argue that the fast excitation of an unsaturated extensile flow is the reason why no nontrivial steady regimes were found in \S~\ref{steady_comp}. This result also highlights the different physical nature of the first two instability modes: self-propelling mode is associated with the symmetry breaking of the advective boundary layer far from the droplet, whereas the second mode encapsulates the interfacial dynamics of the drop.

Linear stability analysis indicates that $\pe = \pe_n$ marks a transition from the isotropic state being linearly-stable ($\pe < \pe_n$) to this base state becoming unstable ($\pe > \pe_n$). For $n=1$, this instability is associated with the onset of self-propulsion discussed in \S~\ref{steady_n1_o2_o3}. Beyond $\pe=\pe_1$, this non-isotropic self-propelled state is itself stable to linear perturbations (see Appendix~\ref{lin2}), and $\pe = \pe_1$ therefore corresponds to an exchange of stability of the two modes as expected for a transcritical bifurcation.

\section{Discussion}
\label{discussion}
In order to elucidate how deformability of chemically active droplets affects the onset of their self-propulsion, the Marangoni instability of an active deformable drop submerged in surfactant solution was analyzed using matched asymptotics expansions near the instability threshold. In this axisymmetric model, the instability is powered by the constant isotropic activity of the droplet (i.e., absorption of surfactant molecules to form swollen micelles) and the advection of the isotropic surfactant concentration field by the fluid motion, while nonlinear dynamics of the model is due to both interface deformations and surfactant advection around the drop.

Two main results were obtained:
\begin{enumerate}[label=(\roman*)]

  \item{deformability was found to enhance self-propulsion of droplets that are more viscous than the surrounding medium (specifically, droplets with viscosity ratio $\eta > 1.0134$), while self-propulsion of less viscous drops ($\eta < 1.0134$) is hindered by the droplet deformability;}

  \item{deformability affects the type of bifurcation leading to symmetry breaking, in particular, moderately deformable droplets exhibit transcritical onset of self-propulsion, while in the case of highly deformable drops our results suggest that the bifurcation becomes subcritical.}

\end{enumerate}
From a physical point of view, the first result (namely the increase of self-propulsion velocity for viscous deformable droplets and the reduction of the velocity for their less viscous counterparts) is the outcome of two different effects associated with the droplet deformation which is always found to generate oblate droplets, namely an increase (resp. decrease) in hydrodynamic drag and reduction (resp. enhancement) in  the front-back surfactant concentration gradient for less (resp. more) viscous droplets.

Investigation of the neutrally stable eigenmodes of the linearized problem further revealed that the interplay between surfactant advection and deformations of the droplet interface results in two competing modes of monotonic instability: the first sets in for the P{\'e}clet number above ${\pe_1 \equiv 4}$ and corresponds to the onset of droplet self-propulsion, whereas the second bifurcates at ${\pe_2 \equiv \left[ 60 ( 1 + \eta ) - \ca ( 4 + \eta ) \right] / ( 2 + 3 \eta )}$ and is characterized by a symmetric extensile flow akin to a flow driven by a force dipole. We argue that these modes reflect two different physical phenomena: the first is associated with the symmetry breaking of the advective boundary layer far from the droplet, whereas the second encapsulates the interfacial dynamics of the drop. The latter is, however, also critically relevant for the self-propulsion as it conditions the hydrodynamic signature of the droplet, its interaction with its neighbors as well as its effect on the macroscopic stress in the fluid~\cite{Batchelor70, Lauga16}.

Above $\pe_1$, the motionless (isotropic) base state of the droplet~\eqref{base} coexists with a regime of finite steady self-propulsion, whose amplitude was determined up to quadratic corrections. The base (motionless) state was observed to lose stability for $\pe > \pe_1$ while the new self-propelled mode is itself stable in that parameter range. Moreover, in the leading order self-propulsion velocity is $\propto \pe - 4$, suggesting that the onset of self-propulsion is a transcritical bifurcation. A similar approach was used to demonstrate that the symmetric steady flow associated with the second transition for $\pe = \pe_2$ can only exist in the case of asymptotically small capillary number, $\ca = O( \epsilon )$, i.e., in the limit of a weakly deformable droplet. Experimentally, it should however not be possible to observe this steady (motionless) state, since in the case of $\ca = O( \epsilon )$, $\pe_1 < \pe_2$, and self-propulsion of the droplet always precedes the onset of a symmetric flow.

Deformability can nevertheless affect self-propulsion itself fundamentally: for highly deformable droplet with ${\ca = 4 ( 13 + 12 \eta ) / ( 4 + \eta ) + O(\epsilon)}$, ${\pe_1 \approx \pe_2}$ and the two instability modes are excited simultaneously. Our results demonstrate that competition between the modes eliminates the regime of steady self-propulsion. On the other hand, we also established that steady extensile flows require $\ca = O( \epsilon )$ and, thus, are also not compatible with high droplet deformability. Consequently, in the case of ${\pe_1 = \pe_2}$, there are no steady flows to be found within the asymptotic limit considered in this paper. This result may be related to an asymptotic disparity in the time scales associated with the first two modes of instability. We presume that unsaturated growth of the second mode hints towards a subcritical nature of the interfacial effects included in the model.

The present work therefore sheds some light on the fundamental role of deformability on the self-propulsion of active droplets. In some recent experimental studies, e.g., Refs.~\cite{Izri14,Moerman17}, the capillary number based on the droplet swimming velocity ($\ca_U=\eta U/\gamma_0$) is typically very small ($\ca_U\sim 10^{-5}$) so that the role of deformability is essentially negligible. Yet, deformability effects can become significant for systems with lower surface tension. For example, spontaneous deformation of chemically active drops due to Marangoni flows was experimentally observed in mm-scale oil drops with an ultra low surface tension of roughly $0.1$~mN/m by Caschera~\textit{et al.}~\cite{Caschera13}, so that $\ca_U\sim 10^{-2}$. In this paper we define the capillary number and dimensionless velocity based on the speed of a drop in an imposed concentration gradient~\eqref{uInf_grad}, since the swimming velocity is not known a priori, resulting in dimensionless terminal velocity~${\sim 0.01}$ (see figure~\ref{propulsion_u}). As a result $\ca \sim 100 \ca_U$ and in an experimental setting, condition~\eqref{def_cond} corresponding to the onset of spontaneous deformations should be met when ${\ca_U \sim 0.1}$.

Asymptotic methods provide significant insight in the interplay of several key physical mechanisms in the dynamics of active droplets, such as surfactant advection by the Marangoni flows or droplet deformation. Consequently, the present study is intrinsically limited to the immediate vicinity of the instability threshold and does not rule out further bifurcations in the dynamical behaviour of such active droplets which require further investigation.

Our findings indicate that the bifurcation structure of steady flows around an active droplet depends on the value of capillary number which quantifies droplet deformability. In that regard, capillary number may be seen as a control parameter: setting the value of $\ca$ determines the nature of the corresponding symmetry breaking bifurcation. We conjecture that the effect of deformability on the dynamics of chemically-driven self-propulsion might be relevant in the context of biology. Indeed, chemically active droplets are widely used to model the behaviour of cells~\cite{Nagasaka17} and it is well established that cells do change their elastic properties dynamically to enhance adhesion and facilitate cell sorting~\cite{Winklbauer15}. Further investigation is thus necessary to elucidate the specific role deformability plays in cell dynamics.

\acknowledgements{
This project has received funding from the European Research Council (ERC) under the European Union's Horizon 2020 research and innovation programme (grant agreement No 714027 to SM).
}

\appendix

\section{Details of the $O(\epsilon^3)$ solution}
\label{details}
In the near field surfactant concentration is given by a superposition of Legendre polynomials, $C^{(3)}( r, \mu ) = \sum\limits_{n=0}^3 C_n^{(3)}(r) L_n(\mu)$, with
\begin{align}
  \label{n1_c3_gen_sol1}
  C_0^{(3)}(r) & = \frac{c_0^{(3)}}{r} + d_0^{(3)}
    - \delta B_1 \frac{8 - 15 r + 20 r^3 + 80 r^6}{120 r^5}
    - \delta^3 \frac{16 - 35 r + 40 r^3 + 160 r^6 - 20 r^7}{30720 r^5}, \\
  \label{n1_c3_gen_sol2}
  C_1^{(3)}(r) & = \frac{c_1^{(3)}}{r^2} + d_1^{(3)} r
    + \delta B_1 \frac{1 + 2 r^3}{2 r^3}
    \nonumber \\ &
    + 3 \delta^3 \frac{
        3 - 7 r + 12 r^2 + 20 r^3 + 28 r^4 + 56 r^6 + 84 r^7 + 56 r^9}
      {143 360 r^7}
    \nonumber \\ & + \frac{33 \delta^3}{627200 r^6 \left( \pe_2 - 4 \right)} 
      \left( 115 - 35 r \frac{24 + 31 \eta}{2 + 3 \eta} 
        + 21 r^2 + 455 r^3 - 210 r^5 \right), \\
  \label{n1_c3_gen_sol3}
  C_2^{(3)}(r) & = \frac{c_2^{(3)}}{r^3} + d_2^{(3)} r^2 
    - \delta B_1 \frac{10 - 21 r + 70 r^3 -42 r^4 + 28 r^6}{84 r^5}
    \nonumber \\ & - \delta^3 \frac{20 - 49 r + 140 r^3 - 112 r^4 + 56 r^6}
      {21504 r^5}
    - 33 \delta^3 \frac{2 + 3 r^2}{1792 r^4 \left( \pe_2 - 4 \right)}, \\
  \label{n1_c3_gen_sol4}
  C_3^{(3)}(r) & = \frac{c_3^{(3)}}{r^4} + d_3^{(3)} r^3 
    + \delta^3 \frac{8 - 21 r + 81 r^2 - 126 r^4 + 162 r^5 
      + 168 r^6 - 63 r^7 + 28 r^9}{215040 r^5}
    \nonumber \\ & + \frac{11 \delta^3}{1756160 r^6 \left( \pe_2 - 4 \right)}
      \bigg( 1274 - 441 r \frac{24 + 31 \eta}{2 + 3 \eta}
        + 216 r^2 + 1512 r^2 \log r 
    \nonumber \\ &
        + 2058 r^3 - 1764 r^4 \frac{9 + 11 \eta}{2 + 3 \eta} + 1764 r^5 \bigg),
\end{align}
where $c_n^{(3)}$ and $d_n^{(3)}$ are unknown constant amplitudes to be determined in the matching process with the far-field boundary layer.

Far field solution writes,
\begin{multline}
  \label{n1_h3_gen_sol}
  H^{(3)}( \rho, \mu )
  = e^{-\rho_s ( 1 + \mu )} \left( 
      -\frac{\delta^3 \rho_s}{192}
      + \delta B_1 \frac{3 + 6 \rho_s - 8 \rho_s^2}{6 \rho_s}
      - \frac{256 \rho_s B_1^2}{3 \delta}
      + 2 b_{o,1}^{(3)} \frac{1 + 2 \rho_s}{\rho_s}
    \right. \\ \left. + \frac{\mu}{256 \delta \rho_s^2} \bigg[
          \delta^4 \left( 3 + 3 \rho_s - 2 \rho_s^3 \right)
        + 128 \delta^2 B_1 \left( 9 + 9 \rho_s + 2 \rho_s^2 - 4 \rho_s^3 \right)
        \right. \\ \left.2
        + 16384 B_1^2 \left( 3 + 3 \rho_s - 2 \rho_s^3 \right)
        + 512 \delta_s b_{o,1}^{(3)} \left( 3 + 3 \rho_s + 2 \rho_s^2 \right)
      \bigg]
    \right. \\ \left. + L_2(\mu) \frac{
        \left( \delta^2 + 128 B_1 \right)^2
          \left( 45 + 45 \rho_s + 15 \rho_s^2 - 2 \rho_s^4 \right)}
        {768 \delta \rho_s^3}
    \right)
  + \frac{ e^{-\rho_s \mu} }{ \sqrt{\rho_s} } 
    \sum\limits_{n=0}^\infty h_n^{(3)} K_{n+1/2} \left( \rho_s \right) L_n(\mu),
\end{multline}
where $h_n^{(3)}$ are unknown constant amplitudes and ${\rho_s \equiv \pe_1 A_1 \rho = \delta \rho / 8 \geq 0}$.

Matching of near- and far-field surfactant concentrations~\eqref{n1_c3_gen_sol1}--\eqref{n1_h3_gen_sol}, and subsequent substitution of the result into the boundary conditions~\eqref{c_consump} and~\eqref{stress_bc}--\eqref{kin_cont} expanded at $O(\epsilon^3)$ yields the solvability condition of the cubic problem providing a correction to the droplet self-propulsion veloicty~\eqref{n1_o3_velocity}.

\section{Linear stability analysis of the steady state featuring self-propulsion}
\label{lin2}
As a complement to the linear stability analysis of the isotropic base state near $\pe_1$ obtained in \S~\ref{steady_n1_o2_o3}, the stability of the non-trivial self-propelled mode obtained for $\pe\geq \pe_1$ is now investigated, in order to demonstrate further that the onset of self-propulsion is a transcritical bifurcation.

As in \S~\ref{lin1}, infinitesimal normal perturbations of the anisotropic state are introduced 
\begin{align}
  \label{lin2_pert1}
  & \psi_i( t, r, \mu )
  = \epsilon \delta \frac{3 r^2 \left( 1 - r^2 \right) 
      \left( 1 - \mu^2 \right)}{64} 
    + e^{\lambda t} \tilde{\psi}_i ( r, \mu ), \\
  & P_i( t, r, \mu )
  = - \epsilon \delta \frac{4 ( 2 + 3 \eta ) + 15 r \mu}{16 \eta} 
    + e^{\lambda t} \tilde{P}_i ( r, \mu ), \\
  & \psi_o( t, r, \mu )
  = \epsilon \delta \frac{\left( 1 - r^3 \right) 
      \left( 1 - \mu^2 \right)}{32 r}
    + e^{\lambda t} \tilde{\psi}_o ( r, \mu ), \\
  & P_o( t, r, \mu )
  = e^{\lambda t} \tilde{P}_i ( r, \mu ), \qquad
  \xi( t, \mu ) = e^{\lambda t} \tilde{\xi} ( \mu ), \\
  \label{lin2_pert3}
  & C( t, r, \mu ) = -\frac{1}{r} 
    + \frac{\epsilon \delta}{8} \left( 1 + \mu 
      + \mu \frac{2 - 3 r}{4 r^3} \right)
    + e^{\lambda t} \tilde{C} ( r, \mu ), \\
  & H( t, \rho, \mu ) = -\frac{\epsilon}{\rho} e^{-\delta \rho ( 1 + \mu ) / 8}
    + e^{\lambda t} \tilde{H} ( \rho, \mu ),
\end{align}
where tilde denotes the perturbations, and $\lambda$ is the perturbations growth rate. Similarly to \S~\ref{lin1}, we focus on the monotonically unstable case, where $\lambda > 0$. Using~\eqref{lin2_pert1}--\eqref{lin2_pert3}, the full nonlinear system~\eqref{shape}--\eqref{forcefree} is linearized around the self-propelling state with respect to perturbations thus obtaining a linear eigenvalue problem for $\lambda$ the associated eigenvector ($\tilde{\psi}_i$, $\tilde{P}_i$, $\tilde{\psi}_o$, $\tilde{P}_o$, $\tilde{C}$, $\tilde{H}$, $\tilde{\xi}$). This problem is tractable within the framework of the matched asymptotic expansions employed in \S~\ref{neutral_modes} and \S~\ref{weakly_nonlin} when $\lambda$ is of the form
\begin{equation}
  \label{lambda_exp}
  \lambda = \epsilon^2 \lambda^{(1)} + \epsilon^3 \lambda^{(2)} + \ldots, 
\end{equation}
and the corresponding eigenvector is expanded as $\tilde{f}=f^{(1)}+\epsilon f^{(2)}+\hdots $, with $f$ being any component of the eigenvector above.
As a result, a sequence of linear problems at ${\epsilon^0, \; \epsilon^1, \; \ldots}$ is obtained, where each of the problems in the sequence represents a Stokes flow past a liquid sphere.

%
%
\subsection{Leading-order problem}
\label{lin2_o0}
The leading order problem is first analyzed. In essence, we apply the algorithm described in \S~\ref{neutral_modes} to the linearized problem at $\epsilon^0$. At $\epsilon^0$, linearized advection-diffusion equation is identical to~\eqref{lin_ad}, and $\tilde{C}^{(1)}$ is therefore given by~\eqref{c1_xi1_modes_raw} and~\eqref{c1_gen_sol1}--\eqref{c1_gen_sol2} -- with potentially different constants $c_n^{(1)}$ and $d_n^{(1)}$ in comparison with \S~\ref{neutral_modes}. The far-field solution is then obtained as,
\begin{equation}
  \label{lin2_h1_gen_sol}
  \tilde{H}^{(1)} \left( \rho, \mu \right) 
  = \frac{ b_{o,1}^{(1)} e^{-\delta \rho ( 1 + \mu ) / 8} }{4 \lambda^{(1)}}
      \left( \frac{\delta}{\rho} + \mu \frac{8 + \delta \rho}{\rho^2} \right) 
  + \frac{e^{-\delta \rho \mu / 8}}{\sqrt{\rho}} \sum\limits_{n=0}^\infty 
      h_n^{(1)} K_{n+1/2} \left( q \rho / 8 \right) L_n(\mu),
\end{equation}
where $q \equiv \sqrt{\delta^2 + 256 \lambda^{(1)}}$. After matching the near- and far-field solutions, we solve the set of algebraic equations emerging form the boundary conditions at the droplet interface. For the self-propelling mode, solution of the leading-order problem for the perturbations reads,
\begin{align}
  \label{lin2_o0_sol1}
  & b_{i,0}^{(1)} = \frac{4 A_1}{3 \eta} ( 2 + 3 \eta ), \qquad
  \frac{3}{2} a_{i,1}^{(1)} = \frac{3}{2} b_{i,1}^{(1)} 
    = a_{o,1}^{(1)} = b_{o,1}^{(1)} = A_1, \\
  \label{lin2_o0_sol2}
  & c_0^{(1)} = 0, \qquad
  d_0^{(1)} = \delta A_1 \frac{q - \delta}{32 \lambda^{(1)}}, \qquad
  c_1^{(1)} = -3 A_1, \qquad
  d_1^{(1)} = 0,
\end{align}
where ${q \equiv \sqrt{\delta^2 + 256 \lambda^{(1)}}}$ and an unknown constant ${A}_1$ will be determined from the solvability condition at $O(\epsilon)$. Solution~\eqref{lin2_o0_sol1}--\eqref{lin2_o0_sol2} is almost identical to the solution of the linearized steady problem~\eqref{o1_sol_n11}--\eqref{o1_sol_n12}, with the exception of the coefficient $d_0^{(1)}$ corresponding to the isotropic perturbation of the surfactant concentration. It is easy to see that in the case of positive perturbation growth rate, ${\lambda > 0}$, coefficient $d_0^{(1)}$ carries the information about the growth rate to the next order of expansion.

%
%
\subsection{Problem at $\epsilon$}
\label{lin2_o1_n1}
Turning now to the $O(\epsilon)$ problem for the perturbations around the self-propelled steady state, we repeat the steps of the analysis developed in \S~\ref{steady_n1_o2_o3}. First, we obtain the solution of the near field advection-diffusion equation, ${\tilde{C}^{(2)}( r, \mu ) = \sum\limits_{n=0}^2 \tilde{C}_n^{(2)}(r) L_n(\mu)}$, where
\begin{align}
  \label{lin2_c2_gen_sol1}
  & \tilde{C}_0^{(2)}(r) = \frac{c_0^{(2)}}{r} + d_0^{(2)} 
  - \delta \tilde{A}_1 \left( \frac{2 r}{3} + \frac{1}{6 r^2} 
    - \frac{1}{8 r^4} + \frac{1}{15 r^5} \right), \\
  & \tilde{C}_1^{(2)}(r) = \frac{c_1^{(2)}}{r^2} + d_1^{(2)} r 
  + \delta \tilde{A}_1 \frac{1 + 2 r^3}{2 r^3} 
  + \frac{2 a_{o,1}^{(2)} + 4 b_{o,1}^{(2)} r^3 }{r^3}, \\
  \label{lin2_c2_gen_sol3}
  & \tilde{C}_2^{(2)}(r) = \frac{c_2^{(2)}}{r^3} + d_2^{(2)} r^2
  + \frac{4 a_{o,2}^{(2)} + 6 b_{o,2}^{(2)} r^2}{r^4}
  - \delta \tilde{A}_1 \left( \frac{r}{3} - \frac{1}{2 r}
    + \frac{5}{6 r^2} - \frac{1}{4 r^4} + \frac{5}{42 r^5} \right).
\end{align}
Then the far-field solution is obtained as,
\begin{multline}
  \label{lin2_h2_gen_sol}
  \tilde{H}^{(2)} \left( \rho, \mu \right) 
  = \frac{ \tilde{A}_1 e^{-\rho \left( q + \delta \mu \right) / 8} }{\lambda^{(1)}} 
      \bigg(
          \delta \frac{ \delta^3 + 160 \delta \lambda^{(1)} 
            + 384 \lambda^{(2)} }{96 q}
        \\ + \frac{\mu}{64} \left( \delta^3 + 64 \delta \lambda^{(1)} 
            + 256 \lambda^{(2)} \right)
        + \delta^2 L_2(\mu) \frac{8 + q \rho}{192 \rho}
      \bigg)
  \\ -\frac{\tilde{A}_1 e^{-\delta \rho ( 1 + \mu ) / 8}}
    {3 \pe_1 \left( \lambda^{(1)} \right)^2} 
      \bigg(
          \delta \frac{\delta^3 + 96 \delta \lambda^{(1)}
            + 384 \pe_1 \lambda^{(2)}}{384 \rho}
        \\ + \mu \left( \delta^3 + 64 \delta \lambda^{(1)} 
            + 256 \lambda^{(2)} \right) \frac{8 + \delta \rho}{256 \rho^2}
        + \delta^2 L_2(\mu) \frac{192 + 24 \delta \rho 
          + \delta^2 \rho^2}{768 \rho^3}
      \bigg)
  \\ + \frac{4 b_{o,1}^{(2)} + \delta \tilde{A}_1}{16 \lambda^{(1)}} 
    e^{-\delta \rho ( 1 + \mu ) / 8}
    \left( \frac{\delta}{\rho} + \mu \frac{8 + \delta \rho}{\rho^2} \right)
  + \frac{e^{-\delta \rho \mu / 8}}{\sqrt{\rho}} \sum\limits_{n=0}^\infty 
      h_n^{(2)} K_{n+1/2} \left( q \rho / 8 \right) L_n(\mu).
\end{multline}
Finally, we match near and far field solutions and use the boundary conditions at the droplet interface~\eqref{c_consump} and~\eqref{stress_bc}--\eqref{kin_cont} for the perturbation fields evaluated up to $O(\epsilon)$ to obtain the solvability condition of the problem at $\epsilon$ reading
\begin{equation} 
  \label{lin2_o1_sol_cond1}
  \ca^2 \left( \pe_1 - \pe_2 \right)
    \left( 2 q^2 - \delta q + 5 \delta^2 \right) = 0.
\end{equation}
It is easy to see that~\eqref{lin2_o1_sol_cond1} is satisfied only when $\ca = O(\epsilon)$. Indeed, limit of $\pe_1 = \pe_2$ is degenerate, since the steady self-propelling regime ceases to exist, while equation~\eqref{lin2_o1_sol_cond1} has no real solution for $q$ above the instability threshold, $\pe > \pe_1$ (recall that self-propelling steady state does not exist for $\pe < \pe_1$).

We repeat the solution of the $O(\epsilon)$ problem for the perturbations of the self-propelled steady state in the limit of a weakly deformable droplet, $\ca = \epsilon \ca_1$. In this case, near and far field solutions remain the same as in the case of finite capillary number, while the solvability condition writes
\begin{equation}
  \label{lin2_o1_sol_cond2}
  \ca_1 \left( q^2 - 2 \delta q + \delta^2 \right) = 0.
\end{equation}
Again, equation~\eqref{lin2_o1_sol_cond2} has no real solutions for $q$ above the instability threshold, $\pe > \pe_1$, and, thus, can be satisfied only when $\ca_1 \ll \epsilon$.

Repeating the solution of the $O(\epsilon)$ problem in the limit of nondeformable droplet, $\ca = 0$, yields near and far field solutions given by~\eqref{lin2_c2_gen_sol1}--\eqref{lin2_h2_gen_sol} and the following solvability condition
\begin{equation}
  \label{lin2_o1_sol_cond3}
  2 q^2 - \delta q + 5 \delta^2 = 0.
\end{equation}
Similarly to the solvability conditions~\eqref{lin2_o1_sol_cond1}--\eqref{lin2_o1_sol_cond2}, equation~\eqref{lin2_o1_sol_cond3} has no real solutions for $q$ above the instability threshold, $\pe > \pe_1$. Finally, we combine solvability conditions~\eqref{lin2_o1_sol_cond1}--\eqref{lin2_o1_sol_cond3} and establish that the perturbation growth rate $\lambda^{(1)}$ is strictly negative and the steady state featuring self-propulsion is stable for $\pe > \pe_1$.

%
%
\bibliographystyle{unsrt}
\bibliography{droplet}

\end{document}